\documentclass[article]{revtex4}
\usepackage[dvips]{graphicx}
\usepackage{float}
\usepackage{array}
\begin{document}

\title{Fluctuation and Order of Antiferromagnetism induced by Paramagnetic Pair-Breaking in Superconducting Vortex Lattice}

\author{Kazushi Aoyama$^{1,2}$ and Ryusuke Ikeda${}^{2}$}

\affiliation{${}^1$ Young Researcher Development Center, Kyoto University, Kyoto 606-8317, Japan\\
${}^2$ Department of Physics, Kyoto University, Kyoto 606-8502, Japan}

\begin{abstract}
Effects of the strong Pauli-paramagnetic pair-breaking (PPB) on the vortex lattice in $d$-wave superconductors are theoretically studied by putting emphasis on consequences of the PPB-induced antiferromagnetic (AFM) ordering in the spatial modulation in the vortex lattice. It is shown that the PPB-induced AFM fluctuation in the superconducting state leads to an enhancement of the vortex lattice form factor which is a measure of spatial variations of the internal magnetic field and that the enhancement becomes more remarkable as an AFM instability is approached. It is also demonstrated that the PPB-induced AFM ordering is assisted by the vortex-lattice modulation, and thus, that the resulting AFM order is spatially modulated, while it is {\it not} localized in the vortex cores but {\it coexistent} with the nonvanishing superconducting order parameter. These results are discussed in connection with two phenomena observed in CeCoIn${}_5$, the anomalous field dependence of the vortex lattice form factor and the AFM order appearing inside the high-field and low-temperature superconducting phase.
\end{abstract}                                                                  

\maketitle

\section{Introduction}
An antiferromagnetic (AFM) instability occurring near a superconducting (SC) phase with $d$-wave symmetry has been repeatedly detected in experiments and discussed theoretically. In zero field, the AFM and SC orders are competitive with each other and tend not to coexist with each other \cite{Ueda}. Recently, AFM quantum critical phenomena near the SC pair-breaking field $H_{c2}(0)$ have been observed commonly in the heavy-fermion superconductors such as CeCoIn${}_5$ \cite{Paglione, Singh, Ronning}, pressured CeRhIn${}_5$ \cite{Park}, NpPd${}_5$Al${}_2$\cite{Honda}, and Ce${}_2$PdIn${}_8$ \cite{Kaczorowski}, which indicates field-tuned proximity of an AFM instability to the onset of the SC order. However, these materials do not show any magnetic order in the normal state: An AFM fluctuation is merely enhanced with decreasing field toward $H_{c2}(0)$ in the nonmagnetic normal state. Further, it has been also clarified in CeCoIn${}_5$ that the AFM fluctuation is enhanced as $H_{c2}(0)$ is approached by increasing the field {\it in the SC phase}. A basic origin of the strong AFM fluctuation induced in the vicinity of $H_{c2}(0)$ seems to consist in characteristic features of these superconductors, i.e., the strong Pauli-paramagnetic pair-breaking (PPB) effect and a {\it d-wave} SC pairing symmetry. In our previous paper \cite{IHA}, we have shown that, in a SC phase with a $d$-wave pairing symmetry, the PPB effect enhanced by increasing field and decreasing temperature tends to induce an AFM order and that the field-induced AFM fluctuation found below $H_{c2}(0)$ can be explained as a result of this novel PPB effect. In this paper, we will extend our previous theory to describe spatial variations of the AFM order in a SC vortex lattice by bearing the experimental studies on CeCoIn${}_5$ in our mind. 
                                                                                                                                                                                                                                                                                                                                                                                                                                                                                                                                                                                                                                                                                                                                                              
CeCoIn${}_5$ is a spin-singlet superconductor with a layered crystal structure along $c$-axis \cite{normal} and its SC pairing symmetry is believed to be of $d_{x^2-y^2}$-type \cite{Hiasa}. 
In the SC state in a magnetic field perpendicular to the basal plane (${\bf H}\parallel c$), an anomalous field dependence of the vortex lattice form factor (VLFF), which is a measure of spatial distribution of the internal magnetic field in vortex lattice states, has been observed in neutron scattering experiments. It has been known that, although VLFF decreases with increasing field in the conventional type II superconductors \cite{VLFF_typeII, Clem}, it rather {\it increases} with increasing field in CeCoIn$_5$, and that the slope of VLFF in this material becomes sharp abruptly in higher fields near $H_{c2}(0)$ at low temperatures \cite{VLFF, White}. First, this intriguing behavior has been explained as a result of strong PPB \cite{IM}. However, the origin of the abrupt increase in higher fields has not been sufficiently explained there. In the preceding theoretical study on VLFF in Ref. \cite{IM}, the strong AFM fluctuation induced in the SC state just {\it below} the $H_{c2}(T)$ curve \cite{Singh, Ronning, mu_SR, Kasahara, Flouquet} is not taken into account. If the AFM order and fluctuation occurring in CeCoIn$_5$ just below $H_{c2}(0)$ are of a SC origin, it is rather necessary to incorporate effects of the PPB-induced AFM fluctuation in explaining VLFF 
data. In addition, a possible relation between such a field-induced increase of VLFF and the proposed FFLO state \cite{Kumagai_06,RI07c} at the high field end of the SC phase should be considered. 
                                                                                                                                                                                                                                                                                                                                                                                                                                                                                                                                                                                                                                                                                                                                                                                                      
On the other hand, in a magnetic field parallel to the basal plane (${\bf H}\parallel ab $), the existence of a high-field and low-temperature (HFLT) SC phase has been clarified previously and has been identified with a spatially 
modulated Fulde-Ferrell-Larkin-Ovchinnikov (FFLO) \cite{FF, LO} vortex lattice state \cite{Bianchi03,Adachi}. However, recent neutron scattering experiments have shown that an AFM order with its staggered moment vector oriented along $c$-axis exists {\it only} inside the HFLT phase \cite{Kenzelmann1, Kenzelmann2}. On the other hand, the doping experiment \cite{Tokiwa} indicating that the HFLT phase is quite sensitive to both the magnetic and nonmagnetic impurities and is destroyed by quite a small amount of nonmagnetic impurities supports the FFLO picture on this HFLT phase \cite{RI_10}. Further, a recent NMR measurement has clarified the presence in the HFLT phase of normal state regions possibly corresponding to the FFLO nodal planes of the SC order parameter, while the AFM order in the HFLT phase is, at least in the higher field region of the HFLT phase, apparently homogeneous in real space \cite{Kumagai_11}. Although several microscopic pictures on an AFM order in the $d$-wave superconductors in high fields have been proposed so far \cite{Yanase, Aperis, Hatakeyama}, these experimental results are comprehensively explained by the scenario that the HFLT phase is a realization of the longitudinal FFLO vortex lattice state \cite{RI_10}, and that the AFM order observed inside the HFLT phase is induced by PPB and favors coexistence with the SC order \cite{Hatakeyama}. 
In Ref. \cite{Hatakeyama}, the spatial distributions of the PPB-induced AFM order in the presence of the FFLO modulation parallel to the applied magnetic field has been theoretically investigated in the Pauli limit where the effect of the in-plane vortex-lattice modulation is neglected. Although the result obtained in the Pauli limit seems to give a correct picture on spatial orderings over larger scales, the vortex lattice modulation must be inevitably included to describe local properties such as the internal magnetic field and the spatial distribution of the AFM order.                                                                                                                                                                                                                                                                   
In this paper, we will investigate the spatial distributions of the internal magnetic field brought by the AFM fluctuation and a possible AFM order in the presence of the vortex lattice, taking account of both the orbital and Pauli-paramagnetic pair-breaking effects in the Ginzburg-Landau (GL) approach where the perturbative expansion with respect to the SC and AFM order parameters is used. It will be shown that VLFF is enhanced by an additional magnetic screening brought by the PPB-induced AFM fluctuation and that the enhancement becomes more remarkable as an AFM instability is approached, which suggests that the anomalous field dependence of VLFF observed in CeCoIn${}_5$ in ${\bf H}\parallel c$ is due to the PPB-induced AFM critical fluctuation. It will also be discussed that the spatial modulation of the SC vortex lattice enhances the AFM fluctuation, and thus that an AFM order with a spatial modulation synchronized with the vortex lattice appears. The modulated AFM is not localized in the vortex core but prefers to coexist with a nonvanishing SC order, which is the same tendency as that of the AFM order modulated by the longitudinal FFLO structure of the SC order parameter \cite{IHA,Hatakeyama}.

The outline of this paper is as follows: In Sec. II, we introduce the theoretical model and derive a Ginzburg-Landau (GL) free-energy functional and a Maxwell equation for the internal magnetic field. The effect of the PPB-induced AFM fluctuation on VLFF in the case with ${\bf H}\parallel c $ is discussed in Sec. III. This is followed by Sec. IV in which we argue the PPB-induced AFM order in the SC vortex lattice state in the case with ${\bf H}\parallel ab$. Summary is provided in Sec. VI.                                
 

\section{Formulation}
\subsection{Electronic Hamiltonian}
We start from the electronic Hamiltonian involving a kinetic energy for noninteracting quasiparticles with the Zeeman energy ${\cal H}_0$ and two interaction terms leading to $d$-wave superconductivity ${\cal H}_{\rm SC}$ and antiferromagnetism ${\cal H}_{\rm AFM }$. It can be written as ${\cal H}={\cal H}_0 + {\cal H}_{\rm SC} + {\cal H}_{\rm AFM}$, where
\begin{eqnarray}\label{eq:H_0}                                                                                                                
{\cal H}_0&=&\sum_{\sigma}\int_{\bf r} \,\varphi^\dagger_\sigma({\bf r}) \bigg[ \varepsilon \Big(-i\, \nabla + |e| {\bf A}({\bf r})\Big)-\sigma \mu_{\rm B} \,g B({\bf r})\bigg] \varphi_\sigma({\bf r}), \nonumber\\
{\cal H}_{\rm SC}&=&-|g|\sum_{\bf q} {\hat \Psi}^\dagger({\bf q}) \, {\hat \Psi}({\bf q}), \nonumber\\
{\cal H}_{\rm AFM }&=&-U \sum_{\bf q} {\hat {\bf S}}^\dagger ({\bf q}) \cdot {\hat {\bf S}} ({\bf q}) 
\end{eqnarray}
with
\begin{eqnarray}
\varphi_\sigma({\bf r})=\frac{1}{\sqrt{V}}\sum_{\bf p} {\hat c}_{{\bf p},\sigma} \, e^{i ({\bf p}_\perp \cdot {\bf r}_\perp + i p_z d \, j)}, \nonumber\\
{\hat \Psi}({\bf q}) =\frac{1}{2}\sum_{{\bf p},\alpha,\beta}  (-i \, \sigma_y)_{\alpha, \beta} \, w_{\bf p} \, {\hat c}_{-{\bf p}+\frac{\bf q}{2},\alpha} {\hat c}_{{\bf p}+\frac{\bf q}{2},\beta}, \nonumber\\
{\hat {\bf S}} ({\bf q})= \sum_{{\bf p},\alpha,\beta} {\hat c}^\dagger_{{\bf p}, \alpha} \, ({\mbox {\boldmath $\sigma$}})_{\alpha,\beta} \, {\hat c}_{{\bf p}+{\bf Q}_0+{\bf q},, \beta}.
\end{eqnarray}                                                                                            
Here, the layered crystal structure with an interlayer distance $d$ in the $z$ direction is assumed, and a position of a quasiparticle ${\bf r}$ is defined by $({\bf r}_\perp, d \, j)$ with an integer $j$ specifying the location of the layer. The unit $\hbar =c= k_{\bf B}=1$ is used throughout this paper. In the Hamiltonian, ${\hat c}_{{\bf p},\alpha}$ is the annihilation operator for a quasiparticle with momentum ${\bf p}$ and spin projection $\alpha$, $\varepsilon({\bf p})$ is a kinetic energy measured from the Fermi level $E_F$, and the Zeeman energy is expressed as $\mu_{\rm B} \,g B({\bf r})$ with the magnetic flux ${\bf B}({\bf r})$, a $g$-factor, and the Bohr magneton $\mu_{\rm B}$. The vector potential ${\bf A}({\bf r})$ is expressed as ${\bf A}({\bf r})={\bf A}_0({\bf r})+{\bf a}({\bf r})$ and is related to the magnetic flux by the equation 
\begin{equation}\label{eq:flux}                                                                                                                                                                                                                                                                                                                                                                 
{\bf B}({\bf r})=\nabla \times {\bf A}({\bf r})=H \, {\hat Z} + \nabla \times {\bf a}({\bf r}),
\end{equation}
where $\nabla \times{\bf A}_0({\bf r})={\bf H}$, and spatially varying internal magnetic fields are given in terms of ${\bf a}({\bf r})$. The direction of the uniform external magnetic field ${\bf H}$ is denoted by ${\hat Z}$ and will be fixed along the $z$ axis (${\hat Z} = {\hat z} $) in ${\bf H}\parallel c$ case and the $y$ axis (${\hat Z}={\hat y}$) in ${\bf H}\parallel ab$ case. Concerning the interaction terms, $|g|$ and $U$ are coupling constants with positive values, $\sigma_i \, (i=x,y,z)$ are the Pauli matrices, $w_{\bf p}$ denotes a SC pairing symmetry, and ${\bf Q}_0=(\pi/a,\pi/a,\pi/d)$ with a lattice constant $a$ in the $ab$ plane is the commensurate nesting vector. Since the pairing symmetry $w_{\bf p}$ is assumed to be of $d_{x^2-y^2}$-type, the identity
\begin{equation}                                                               
w_{{\bf p}+{\bf Q}_0}=-w_{\bf p}
\end{equation}
is satisfied. In our model Hamiltonian, antiferromagnetism is induced by the nesting property of the dispersion                                                        
\begin{equation}\label{eq:dispersion_real}
\varepsilon({{\bf p}+{\bf Q}_0})=-\varepsilon({\bf p})+T_c \delta_{\rm IC} ,
\end{equation}
where the deviation from the perfect nesting condition is scaled by the SC transition temperature $T_c$ and is measured by the dimensionless parameter $\delta_{\rm IC}$ \cite{Chubukov}. Although $\delta_{\rm IC}$ is ${\bf p}$-dependent in general, $\delta_{\rm IC}$ is assumed to be a constant value since the details of the dispersion do not change our result qualitatively \cite{Hatakeyama}. Then, the Fermi velocity vector ${\bf v}_{\bf p}$ defined by $d \, \varepsilon({\bf p})/d \, {\bf p}$ satisfies the relation ${\bf v}_{{\bf p}+{\bf Q}_0} = -{\bf v}_{\bf p}$.
                                                             
In order to discuss AFM order and fluctuation in the SC state, we introduce the SC pair-field $\Delta({\bf q})$ and the AFM staggered field ${\bf m}({\bf q})$ which are defined by
\begin{eqnarray}                                                                                                                                                                                                                                                                                                        
\Delta({\bf q})&=&|g| \langle {\hat \Psi}({\bf q}) \rangle ,\nonumber\\
{\bf m}({\bf q})&=&U \langle {\hat {\bf S}}({\bf q}) \rangle,
\end{eqnarray}
where $\langle \, \rangle$ represents the statistical average. Here, $\Delta({\bf q})$ (${\bf m}({\bf q})$) plays a role of the SC (AFM) order parameter. In the mean-field approximation, the interaction terms in the Hamiltonian are expressed as 
\begin{eqnarray}\label{eq:H_SCAF}                           
{\cal H}_{\rm SC} &=&  \frac{1}{|g|}\sum_{\bf q}|\Delta({\bf q})|^2-\sum_{\bf q}\Big(  \Delta({\bf q}) \, {\hat \Psi}^\dagger({\bf q}) + {\rm H.c.} \Big),  \nonumber\\
{\cal H}_{\rm AFM} &=&  \frac{1}{U}\sum_{\bf q}|{\bf m}({\bf q})|^2- \sum_{\bf q} \Big( {\bf m}({\bf q}) \cdot {\hat {\bf S}}^\dagger({\bf q}) + {\rm H.c.} \Big).
\end{eqnarray}
Although, in principle, the AFM moment vector ${\bf m}({\bf q})$ can be oriented in any direction, the two typical configurations, ${\bf m}\parallel {\bf H}$ and ${\bf m} \perp {\bf H}$, will be considered since, in CeCoIn${}_5$ of our interest, the configuration ${\bf m}\parallel c \perp {\bf H} $ has been confirmed in the AFM order observed inside the HFLT phase in ${\bf H}\parallel ab$ case and ${\bf m}\parallel{\bf H}$ would be realized in ${\bf H}\parallel c$ case if ${\bf m}$ is locked in the $c$-axis. 

\subsection{Ginzburg-Landau free energy functional}
The free energy of this system is given by $F=-T \ln {\rm Tr}_{c,c^\dagger,\Delta,\Delta^*,m}\big( \exp\big[-({\cal H}_0+{\cal H}_{\rm SC}+{\cal H}_{\rm AFM })/T\big] \big) + \int d^3r {\bf B}^2({\bf r})/(8\pi)$. In this paper, we derive the Ginzburg-Landau (GL) free energy of a form expanded in powers of both $|\Delta({\bf q})|$ and $|m({\bf q})|$, taking both the orbital and Pauli-paramagnetic pair-breaking effects into account. Formally, the mean-field GL free-energy density in the present system can be written as
\begin{equation}
{\cal F}_{\rm GL}(\Delta,m)={\cal F}^{(2)}_\Delta+{\cal F}^{(4)}_\Delta+{\cal F}^{(6)}_\Delta + {\cal F}^{(2)}_m+{\cal F}^{(2,2)}_{\Delta,m}+{\cal F}^{(4)}_m,
\end{equation}                        
where ${\cal F}^{(l)}_{\Delta}$ (${\cal F}^{(l)}_{m}$) is the $l$th order term dependent only on $|\Delta|$ ($|m|$), and the leading order SC-AFM coupling term ${\cal F}^{(2,2)}_{\Delta, m }$, which is proportional to $|\Delta|^2 \, |m|^2$, is incorporated. We note that, although the correction to the $|m|^4$ term ${\cal F}^{(2,4)}_{\Delta,m}$, which is proportional to $|\Delta|^2|m|^4$, should be also incorporated in examining the character of the AFM transition, ${\cal F}^{(2,4)}_{\Delta,m}$ is omitted here since we have already checked that it tends to make the AFM transition a continuous one \cite{Hatakeyama}. First, we determine the $H_{c2}(T)$ curve and the amplitude of the spatially averaged gap function $\Delta^2\equiv\langle |\Delta({\bf r})|^2\rangle_{sp}$ from the SC part of the GL free energy ${\cal F}^{(2)}_\Delta+{\cal F}^{(4)}_\Delta+{\cal F}^{(6)}_\Delta$. This procedure in which $\Delta$ and $m$ are separately considered is justified at least near the second order AFM transition. Since ${\cal F}_{\rm GL}(\Delta,m=0)$ takes the form
\begin{equation}\label{eq:F_s}                                                                                                                                                                                                                                                                                                                                                                                                        
{\cal F}_{\rm GL}(\Delta,0)=V_2 |\Delta|^2+\frac{V_4}{2}|\Delta|^4+\frac{V_6}{3}|\Delta|^6
\end{equation}
with the coefficients $V_i$ whose expressions reflect the microscopic details, the discontinuous $H_{c2}(T)$ curve is determined by
\begin{equation}
V_2=\frac{3}{16} \frac{V_4^2}{V_6}
\end{equation}
and the $\Delta$ which minimizes ${\cal F}_{\rm GL}(\Delta,m=0)$ is determined by
\begin{equation}
|\Delta_0|^2=\frac{-V_4+\sqrt{V_4^2 - 4 V_2 V_6}}{2V_6}.
\end{equation}
Next, AFM fluctuation inside the SC phase determined by ${\cal F}_{\rm GL}(\Delta,0)$ will be discussed based on the usual GL theory for the AFM part of the free energy ${\cal F}^{(2)}_m+{\cal F}^{(2,2)}_{\Delta_0,m}+{\cal F}^{(4)}_m$. 
The AFM instability is determined by 
\begin{equation}\label{eq:AF_inst}
\frac{1}{m^2} \Big({\cal F}^{(2)}_m+{\cal F}^{(2,2)}_{\Delta_0,m} \Big) =0,
\end{equation}
where $m^2 \equiv \langle |m({\bf r})|^2\rangle_{sp}$ is the spatially averaged value of the AFM order parameter. The $m$ minimizing ${\cal F}_{\rm GL}(\Delta_0,m)-{\cal F}_{\rm GL}(\Delta_0,0)$ is determined by 
\begin{equation} \label{eq:AFM_amp}
|m_0|^2=-\frac{1}{2}\frac{\big({\cal F}^{(2)}_m+{\cal F}^{(2,2)}_{\Delta_0,m}\big)/m^2}{{\cal F}^{(4)}_m \, /\, m^4}.
\end{equation}

To derive the GL free-energy functional ${\cal F}_{\rm GL}$, we will carry out the Feynman-diagrammatic calculation. Regarding the field dependence of the quasiparticle Green's function $-\langle T_\tau \varphi_\sigma({\bf r},\tau) \, \varphi^\dagger_{\sigma} ({\bf r}',0) \rangle = T\sum_{\varepsilon_n} G_{\varepsilon_n, \sigma}({\bf r},{\bf r'}) \,e^{-\varepsilon_n \tau}$ with a fermion Matsubara frequency $\varepsilon_n=\pi T (2n+1)$, we will use the quasi-classical approximation for the Green's function defined in the normal state ${\cal G}_{\varepsilon_n, \sigma}({\bf r},{\bf r}_1)$, namely,
\begin{equation}\label{eq:quasiclassicalapp}                                                                                                                                                                                                                                                                                                                                                                                                                                                                                                                                                                                                                                                                                     
{\cal G}_{\varepsilon_n, \sigma}({\bf r},{\bf r}_1) \simeq {\cal G}_{\varepsilon_n, \sigma}({\bf r}-{\bf r}_1) \, e^{i |e| \int_{{\bf r}}^{{\bf r}_1} d{\bf s}\cdot {\bf A}({\bf s})},
\end{equation}
where ${\cal G}_{\varepsilon_n, \sigma}({\bf r}-{\bf r}_1)$ is defined in the uniform normal state and its Fourier transformation is given by
\begin{equation}
{\cal G}_{\varepsilon_n, \sigma}({\bf p})=\big(i\varepsilon_n -\big[ \, \varepsilon({\bf p}) -\sigma \,\overline{I} T_c  \big] \big)^{-1} 
\end{equation}
with $\overline{I}=\mu_{\rm B}\, g H /T_c$. The orbital pair-breaking effect, which is brought by $\exp [i |e| \int d{\bf s}\cdot {\bf A}({\bf s})]$ in Eq. (\ref{eq:quasiclassicalapp}), can be fully incorporated in the diagrammatic calculation by using the relation \cite{Werthamer}                                                                                                                                                       
\begin{equation}                                
{\rm exp}\Big( i 2|e| \int_{\bf r}^{{\bf r}_1} d{\bf s}\cdot {\bf A}({\bf s})\Big) \Delta({\bf r}_1)=e^{-i ({\bf r}-{\bf r}_1)\cdot {\bf \Pi}({\bf r})}\Delta({\bf r}) 
\end{equation}
with
\begin{equation}
{\bf \Pi}({\bf r}) = -i \, \nabla +2|e|{\bf A}({\bf r}).
\end{equation}  

\begin{figure}[t]
\begin{center}
\includegraphics[scale=0.4]{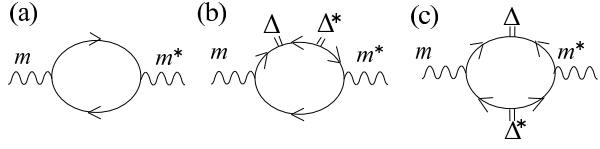}
\caption{Feynman diagrams describing the second order terms with respect to the AFM order parameter $|m|$ in the GL free energy. The $|m|^2$ term in the normal state ${\cal F}_m^{(2)}$ is described by (a) and the correction term brought by the SC order ${\cal F}_{\Delta,m}^{(2,2)}$ consists of the contributions coming from the two types of diagrams (b) and (c). A solid line denotes the quasiparticle Green's function defined in the normal state. \label{fig:Feynman_AF}}
\end{center}                                                                                                                                                                                                                                    
\end{figure}

The quadratic, quartic, and sixth order terms with respect to $|\Delta|$ have been already derived elsewhere \cite{Adachi, RI07ab} and are written as 
\begin{eqnarray}
{\cal F}_\Delta^{(2)}&=&\frac{1}{V}\int_{\bf r} \, \Delta^\ast({\bf r})\Big( \frac{1}{|g|}-\hat{K}_\Delta^{(2)}({\bf \Pi}) \Big)\Delta({\bf r}), \nonumber\\
\hat{K}_\Delta^{(2)}({\bf \Pi})&=& \frac{T}{2}\sum_{\varepsilon_n,\sigma} \sum_{\bf p} \, |w_{\bf p}|^2 \, {\cal G}_{\varepsilon_n, \sigma}({\bf p}) \, {\cal G}_{-\varepsilon_n, -\sigma}(-{\bf p}+{\bf \Pi}), \nonumber\\
{\cal F}_\Delta^{(4)}&=&\frac{1}{2V}\int_{\bf r} \, \hat{K}_\Delta^{(4)}({\bf \Pi}_i) \Delta^\ast({\bf s}_1)\Delta({\bf s}_2) \Delta^\ast({\bf s}_3)\Delta({\bf s}_4) \big|_{{\bf s}_i\rightarrow{\bf r}}, \nonumber\\
\hat{K}_\Delta^{(4)}({\bf \Pi}_i)&=& \frac{T}{2}\sum_{\varepsilon_n,\sigma} \sum_{\bf p} \, |w_{\bf p}|^4 \, {\cal G}_{\varepsilon_n, \sigma}({\bf p}) \, {\cal G}_{-\varepsilon_n, -\sigma}(-{\bf p}+{\bf \Pi}^\dagger_1) \,  {\cal G}_{-\varepsilon_n, -\sigma}(-{\bf p}+{\bf \Pi}_2) \, {\cal G}_{\varepsilon_n, \sigma}({\bf p}+{\bf \Pi}^\dagger_3-{\bf \Pi}_2),\nonumber\\
{\cal F}_\Delta^{(6)}&=&\frac{1}{3V}\int_{\bf r} \, \hat{K}_\Delta^{(6)}({\bf \Pi}_i) \Delta^\ast({\bf s}_1)\Delta({\bf s}_2) \Delta^\ast({\bf s}_3)\Delta({\bf s}_4) \Delta^\ast({\bf s}_5)\Delta({\bf s}_6)\big|_{{\bf s}_i\rightarrow{\bf r}}, \nonumber\\                                           
\hat{K}_\Delta^{(6)}({\bf \Pi}_i)&=& -\frac{T}{2}\sum_{\varepsilon_n,\sigma} \sum_{\bf p} \, |w_{\bf p}|^6 \, {\cal G}_{\varepsilon_n, \sigma}({\bf p}) \, {\cal G}_{-\varepsilon_n, -\sigma}(-{\bf p}+{\bf \Pi}^\dagger_1) \, {\cal G}_{-\varepsilon_n, -\sigma}(-{\bf p}+{\bf \Pi}_6) \nonumber\\
&&\times {\cal G}_{\varepsilon_n, \sigma}({\bf p}-{\bf \Pi}^\dagger_1+{\bf \Pi}_2) \, {\cal G}_{-\varepsilon_n, -\sigma}(-{\bf p}+{\bf \Pi}^\dagger_1-{\bf \Pi}_2+{\bf \Pi}^\dagger_3) \, {\cal G}_{\varepsilon_n, \sigma}({\bf p}-{\bf \Pi}_6+{\bf \Pi}^\dagger_5).
\end{eqnarray}         
The concrete expressions of the coefficients $V_i$ in Eq. (\ref{eq:F_s}) are given in Appendix II. 
The AFM contributions in the GL free energy are formally written as
\begin{eqnarray}\label{eq:F_af}
{\cal F}^{(2)}_m&=& \sum_{\bf q} \Big( \frac{1}{U}+\hat{K}_m^{(2)}({\bf q})\Big) \, |m({\bf q})|^2, \nonumber\\
{\cal F}^{(2,2)}_{\Delta,m}&=& \sum_{{\bf q},{\bf q}'} \hat{K}_{\Delta,m}^{(2,2)}({\bf q},{\bf q}')  \, m({\bf q})m^\ast({\bf q}'), \nonumber\\
{\cal F}^{(4)}_m&=& \frac{1}{2}\sum_{{\bf q}_i \, (i=1-4)} \hat{K}_m^{(4)}({\bf q}_i) \, m({\bf q}_1)m^\ast({\bf q}_2)m({\bf q}_3)m^\ast({\bf q}_4).
\end{eqnarray}                                          
The $|m|^2$ terms are described by Feynman diagrams shown in Fig. \ref{fig:Feynman_AF}. Figure \ref{fig:Feynman_AF} (a) corresponds to ${\cal F}^{(2)}_m$, and ${\cal F}^{(2,2)}_{\Delta,m}$ consists of the contributions described by the two types of diagrams (b) and (c) in Fig. \ref{fig:Feynman_AF}. Then, $ \hat{K}_m^{(2)}({\bf q})$ and $\hat{K}_{\Delta,m}^{(2,2)}({\bf q},{\bf q}')$ are given by 
                                                                                                                                                                                                                                                                                   

\begin{eqnarray}\label{eq:AF2nd}
\frac{1}{U}+\hat{K}_m^{(2)}({\bf q})&=& \bigg( \frac{1}{U}+\frac{1}{2}\sum_\sigma T\sum_{\varepsilon_n} \sum_{\bf p} {\cal G}_{\varepsilon_n,\sigma}({\bf p}){\cal G}_{\varepsilon_n,{\overline \sigma}}({\bf p}+{\bf Q}_0+{\bf q}) \bigg) \nonumber\\
&=&N(0) \bigg( \ln \frac{T}{T_{\rm N}} + 2\pi T\sum_{\varepsilon_n>0} \bigg[ \frac{1}{|\varepsilon_n|}-\frac{1}{2}\sum_\sigma \sum_{s_{\varepsilon_n} = \pm 1} \, \Big\langle \frac{i\, s_{\varepsilon_n}  }{d_{+,q}^{(-)} }\Big\rangle_{\rm FS} \bigg] \bigg), \nonumber\\
\hat{K}_{\Delta,m}^{(2,2)}({\bf q},{\bf q}')&=& -\frac{1}{2}\sum_\sigma T\sum_{\varepsilon_n} \sum_{\bf p} \frac{1}{V}\int_{\bf r} \, \bigg[ 2\,|w_{\bf p}|^2 \, {\cal G}_{\varepsilon_n,\sigma}({\bf p}) \, {\cal G}_{\varepsilon_n,{\overline \sigma}}({\bf p}-{\bf Q}_0-{\bf q}) \, {\cal G}_{\varepsilon_n,\sigma}({\bf p}-({\bf q}-{\bf q}')) \, {\cal G}_{-\varepsilon_n,-\sigma}(-{\bf p}+{\bf \Pi}_1+({\bf q}-{\bf q}'))\nonumber\\
&& - \,w_{\bf p}w_{{\bf p}+{\bf Q}_0} \, {\cal G}_{\varepsilon_n,\sigma}({\bf p}) \, {\cal G}_{\varepsilon_n,{\overline \sigma}}({\bf p}-{\bf Q}_0-{\bf q}) \, {\cal G}_{-\varepsilon_n,-{\overline \sigma}}(-{\bf p}+{\bf Q}_0+{\bf q}+{\bf \Pi}_1) \, {\cal G}_{-\varepsilon_n,-\sigma}(-{\bf p}+{\bf \Pi}_1+({\bf q}-{\bf q}')) \bigg]  \nonumber\\
&& \times e^{i \, {\bf r}\cdot({\bf q}-{\bf q}')} \Delta^\ast({\bf r}) \, \Delta({\bf s}_1)|_{{\bf s}_1 \rightarrow {\bf r}} \nonumber\\
&=&-\frac{1}{2}\sum_\sigma 2\pi T N(0) \sum_{\varepsilon_n} (i \,s_{\varepsilon_n}) \int_{\bf r} \, e^{i \, {\bf r}\cdot({\bf q}-{\bf q}')} \Delta^\ast({\bf r})\bigg\langle 2\,|w_{\bf p}|^2 \bigg[\frac{1}{d_{\sigma,\Pi_1}^{(-)} \, d_{\sigma,\Pi_1+q-q'}^{(-)}  \, d_{+,q}^{(-)}} \nonumber\\
&&+\frac{1}{d_{\sigma,\Pi_1}^{(-)} \, d_{+,q}^{(-)} \, d_{+,q'}^{(-)}}\bigg]- w_{\bf p}w_{{\bf p}+{\bf Q}_0} \bigg[ -\frac{1}{d_{\sigma,\Pi_1+q-q'}^{(-)} \, d_{+,q}^{(-)} \, d_{+,q'}^{(+)}} -\frac{1}{d_{\overline {\sigma},\Pi_1}^{(+)} \, d_{+,q}^{(-)} \, d_{+,q'}^{(+)} }\bigg] \, \Delta({\bf s}_1)\bigg\rangle_{\rm FS}  \big|_{{\bf s}_1 \rightarrow {\bf r}} \nonumber\\
\end{eqnarray}
with
\begin{eqnarray}
d_{\sigma,\Pi_i}^{(\pm)}&=&2i\, \varepsilon_n + 2\sigma \overline{I} T_c \pm {\bf v}_{\bf p}\cdot {\bf \Pi}_i \, ,\nonumber\\
d_{+,q}^{(\pm)}&=&2i\, \varepsilon_n + (\sigma + {\overline \sigma}) \overline{I} T_c \pm (T_c\delta_{\rm IC}+{\bf v}_{\bf p}\cdot {\bf q}),
\end{eqnarray}
where $s_{\varepsilon_n}$ represents ${\rm sgn}(\varepsilon_n)$, the summation for the momentum ${\bf p}$ has been carried out by using the usual replacement $\sum_{\bf p} \rightarrow N(0)\int d \varepsilon({\bf p})  \, \big\langle \big\rangle_{\rm FS}$, $N(0)$ is the density of state per spin at the Fermi level, $\big\langle  \big\rangle_{\rm FS}$ represents the angle average on the Fermi surface, and the approximation $\varepsilon({\bf p}+{\bf \Pi}) \simeq \varepsilon({\bf p})+{\bf v}_{\bf p}\cdot {\bf \Pi}$ ($|{\bf \Pi}|/|{\bf p}| \ll 1$) is used. $T_N$ is the AFM transition temperature in the normal state. It should be noted that ${\overline \sigma}$ changes its sign depending on the orientation of ${\bf m}$ relative to the external magnetic field ${\bf H}$, i.e., ${\overline \sigma} = \sigma$ for $ {\bf m} \parallel {\bf H} $ and  ${\overline \sigma} = -\sigma$ for ${\bf m} \perp {\bf H} $.

By using the transformation 
\begin{equation}\label{eq:parameter_int}
\frac{1}{D} = \int_0^{\infty} d\rho \, \exp\Big[-D \, \rho \Big] \quad ({\rm Re} \, D > 0),
\end{equation}
Eq. (\ref{eq:AF2nd}) can be rewritten as
\begin{equation}
\frac{1}{U}+\hat{K}_m^{(2)}({\bf q})= N(0) \Big( \ln \frac{T}{T_{N}} + \int_0^\infty d\rho \bigg[ \frac{2\pi t}{\sinh (2\pi t \, \rho)}-f_{\cos} (\rho,0) \Big\langle \cos\Big((\delta_{\rm IC} + T_c^{-1}{\bf v}_{\bf p}\cdot{\bf q}) \, \rho \Big)\Big\rangle_{\rm FS} \bigg]\Big) 
\end{equation}
and                                                                                                                                                                                                                                                                                                      
\begin{eqnarray}\label{eq:SC2AF_0}
\hat{K}_{\Delta,m}^{(2,2)}({\bf q},{\bf q}')&=& \frac{N(0)}{T_c^2} \prod_{i=1}^3 \int_0^\infty d\rho_i  \frac{1}{V}\int_{\bf r} \,e^{i \, {\bf r}\cdot({\bf q}-{\bf q}')} \Delta^\ast({\bf r}) \frac{1}{2}\sum_{s_{\varepsilon_n}=\pm 1}\sum_{j=1}^4 P_j \, f_{\cos} \big(\sum_{i=1}^3 \rho_i, \, A_j \big)  \nonumber\\
&&\times \bigg\langle \,|w_{\bf p}|^2  \, {\rm exp}\Big[i s_{\varepsilon_n} (\delta_{\rm IC} (B_j + C_j) + \frac{{\bf v}_{\bf p}\cdot {\bf q}}{T_c}B_j+\frac{{\bf v}_{\bf p}\cdot {\bf q}'}{T_c} C_j \Big] \exp\Big[i s_\varepsilon T_c^{-1}{\bf v}_{\bf p}\cdot {\bf \Pi}_1 \, A_j \Big]  \Delta({\bf s}_1) \bigg\rangle_{\rm FS}\Big|_{{\bf s}_1 \rightarrow {\bf r}},
\end{eqnarray} 
where the relation $w_{{\bf p}+{\bf Q}_0}=-w_{\bf p}$ is used, the functions 
\begin{eqnarray}\label{eq:f}
f_{\cos} (x,y) &=&\frac{2\pi t}{\sinh\big[2\pi t \,x\big]} \cos\Big(\overline{I} \, \Big[2 \, y +\frac{\sigma+\overline{\sigma}}{\sigma}(x-y)\Big]\Big), \nonumber\\
f_{\sin} (x,y) &=&\frac{2\pi t}{\sinh\big[2\pi t \,x\big]} \sin\Big(\overline{I} \, \Big[2 \, y +\frac{\sigma+\overline{\sigma}}{\sigma}(x-y)\Big]\Big)
\end{eqnarray}
are introduced for convenience, and the coefficients $P_j$, $A_j$, $B_j$, and $C_j$ are shown in Table. \ref{table:AF2nd}. In the same manner, $\hat{K}_m^{(4)}({\bf q}_i)$ in Eq. (\ref{eq:F_af}) is calculated as 
\begin{eqnarray}
\hat{K}_m^{(4)}({\bf q}_i)&=&T\sum_{\varepsilon_n,\sigma} \sum_{\bf p}  \, {\cal G}_{\varepsilon_n,\sigma} ({\bf p}) \, {\cal G}_{\varepsilon_n,\overline{\sigma}} ({\bf p}+{\bf Q}_0+{\bf q}_2) \, {\cal G}_{\varepsilon_n,\sigma} ({\bf p}+{\bf q}_2-{\bf q}_3) \, {\cal G}_{\varepsilon_n,\overline{\sigma}} ({\bf p}+{\bf Q}_0+{\bf q}_1) \, \delta_{\Sigma_q}\nonumber\\
&=&\delta_{\Sigma_q}\,\frac{N(0)}{T_c^2}\prod_{i=1}^3\int_0^\infty d\rho_i \,f_{\cos} \big(\sum_{i=1}^3 \rho_i,0 \big) \, \cos\Big(\delta_{\rm IC} \, \big(\sum_{i=1}^3 \rho_i \big)\Big) \nonumber\\
&& \times \bigg\langle \cos\Big(T_c^{-1}\Big[ {\bf v}_{\bf p}\cdot{\bf q}_1 (\rho_1+\rho_2)+{\bf v}_{\bf p}\cdot{\bf q}_2 (-\rho_2+\rho_3) + {\bf v}_{\bf p}\cdot{\bf q}_3 \rho_2 \Big]\Big) \nonumber\\
&& + \cos\Big(T_c^{-1}\Big[ {\bf v}_{\bf p}\cdot{\bf q}_1 \rho_1 + {\bf v}_{\bf p}\cdot{\bf q}_2 (-\rho_1+\rho_2) + {\bf v}_{\bf p}\cdot{\bf q}_3 (\rho_1+\rho_3) \Big]\Big)\bigg\rangle_{\rm FS},
\end{eqnarray}   
where $\delta_{\Sigma_q}$ denotes $\delta_{q_4,q_1-q_2+q_3}$. 

\begin{table}[t]
\caption{Coefficients $P_j$, $A_j$, $B_j$, and $C_j$ in Eqs. (\ref{eq:SC2AF_0}) and (\ref{eq:SC2AF})\label{table:AF2nd}}
\begin{tabular}{>{$}c<{$}>{$}c<{$}>{$}c<{$}>{$}c<{$}>{$}c<{$}}
\hline
\hline
j & P_j & A_j & B_j & C_j  \\
\hline
1 & 2  & \rho_1+\rho_2 &  \rho_2+\rho_3 & -\rho_2       \\ 
2 & 2  & \rho_1        &  \rho_3        & \rho_2        \\
3 & -1 & \rho_1        &  \rho_1+\rho_3 & -\rho_1-\rho_2\\
4 & -1 & \rho_1        &  -\rho_3       & \rho_2        \\
\hline
\hline
\end{tabular}
\end{table}

Next, we calculate $\int_{\bf r} \,e^{i \, {\bf r}\cdot({\bf q}-{\bf q}')} \Delta^\ast({\bf r}) \, \exp \big[ i s_\varepsilon A \, T_c^{-1}{\bf v}_{\bf p}\cdot {\bf \Pi}_i \big] \Delta({\bf s}_i)|_{{\bf s}_i \rightarrow {\bf r}}$ in Eq. (\ref{eq:SC2AF_0}). In the presence of a magnetic field, the SC gap function $\Delta({\bf r})$ has a spatial modulation due to vortices induced by the orbital pair-breaking effect. In this paper, the SC gap function is assumed to take the form of a familiar Abrikosov vortex lattice 
\begin{equation}
\Delta({\bf r})=\Delta \varphi_0(X,Y)
\end{equation}
with 
\begin{equation}
\varphi_0(X,Y)={\sqrt \frac{k}{\sqrt \pi}}\sum_{s=-\infty}^\infty \exp \bigg[ i \, \Big( \frac{sk}{r_H}Y + \frac{\pi}{2}s^2\Big)-\frac{1}{2}\Big(\frac{X}{r_H}+sk\Big)^2\bigg],
\end{equation}
where $r_H$ is the magnetic length defined by $r_H=(2|e|H)^{-1/2}$ and a structure of the vortex lattice is specified by $k$. The coordinate $(X,Y)$ denotes that in the plane perpendicular to the magnetic field and is defined by $(x,y)$ for ${\hat Z}\parallel{\hat z}$ and $(\gamma^{1/2}z,\gamma^{-1/2}x)$ for ${\hat Z}\parallel {\hat y}$, where $\gamma$ denotes the ratio of the SC coherence length in the basal ($ab$) plane to that in the direction along $c$-axis and is expressed as $\sqrt{ \langle v_{{\bf p},x}^2 \rangle _{\rm FS} /\langle v_{{\bf p},z}^2 \rangle _{\rm FS}}$. In our numerical calculation, $k=\sqrt{\pi \sqrt{3}}$ describing the triangular lattice is used since, in the high field SC phase of CeCoIn${}_5$ in ${\bf H}\parallel c$, the square lattice characteristic of $d$-wave superconductors is deformed into nearly triangular ones due to the strong PPB effect \cite{VLFF, Hiasa}. By using the identity \cite{Adachi}
\begin{equation}\label{eq:orbital_operation}
\exp\Big[i A \frac{{\bf v}_{\bf p}\cdot {\bf \Pi}}{T_c}\Big]\varphi_0(X,Y)={\rm exp} \Big[ -\frac{1}{2} \, A^2 \, (|\eta |^2 -\eta^{\ast 2})\Big] \varphi_0(X+ A\,  \eta^\ast {\sqrt 2} r_H,Y) 
\end{equation}
with $\eta= \big( v_{{\bf p},x} + i \, v_{{\bf p},y}\big)/ \big({\sqrt 2} \, r_H \, T_c\big)$ for ${\hat Z}\parallel{\hat z}$ and $\eta=\big( \gamma^{1/2} v_{{\bf p},z} + i \, \gamma^{-1/2}v_{{\bf p},x}\big)/\big({\sqrt 2} \, r_H \, T_c \big)$ for ${\hat Z}\parallel {\hat y}$, we obtain
\begin{eqnarray}\label{eq:SC2AF}
\hat{K}_{\Delta,m}^{(2,2)}({\bf q},{\bf q}')&=& N(0) \frac{|\Delta|^2}{T_c^2} \prod_{i=1}^3 \int_0^\infty d\rho_i  \frac{1}{V}\int_{\bf r} e^{i({\bf q}-{\bf q}')\cdot{\bf r}} \, \sum_{{\bf K}_\perp} T^{(2)}_{{\bf K}_\perp} \, e^{i{\bf K}_\perp\cdot(X,Y)} \sum_{j=1}^4 P_j \, f_{\cos} \big(\sum_{i=1}^3 \rho_i, \, A_j \big) \cos\Big(\delta_{\rm IC} \, (B_j+C_j)\Big) \nonumber\\
&&\times \bigg\langle |w_{\bf p}|^2 \exp\Big[-\frac{|\eta|^2}{2} \, A_j^2 \Big]  \cos\Big(\frac{{\mbox{\boldmath $\eta$}}\cdot {\bf K}_\perp r_H }{{\sqrt 2}} A_j + \frac{{\bf v}_{\bf p}\cdot{\bf q}}{T_c} B_j + \frac{{\bf v}_{\bf p}\cdot{\bf q}'}{T_c}C_j\Big) \cosh\Big( \frac{({\mbox{\boldmath $\eta$}}\times {\bf K}_\perp r_H )_z}{{\sqrt 2}} A_j \Big)  \bigg\rangle_{\rm FS},
\end{eqnarray}
where ${\mbox{\boldmath $\eta$}}=({\rm Re} \, \eta,{\rm Im} \, \eta)$ and ${\bf K}_\perp=(K_x,K_y)$ is a reciprocal lattice vector of the SC vortex lattice in the plane perpendicular to the magnetic field. The details of the calculation in incorporating the orbital pair-breaking effect and the useful identities which will be used below are shown in Appendix I. 

According to Eq. (\ref{eq:SC2AF}), the SC-AFM coupling term is nonvanishing only when 
$(q_x-q'_x, \, q_y-q'_y, \, q_z-q'_z)$ is equal to  $ -(K_x,K_y,0) $ in ${\hat Z}\parallel{\hat z}$ and $-(\gamma^{-1/2}K_y,0,\gamma^{1/2} K_x) $ in ${\hat Z}\parallel {\hat y}$.

\subsection{Vortex lattice form factor and the internal magnetic field}
In the vortex lattice state, the SC gap function is expressed as $|\Delta(X,Y)|^2=\sum_{{\bf K}_\perp} T^{(2)}_{{\bf K}_\perp} \exp\Big[i \, {\bf K}_\perp \cdot \,(X,Y)\Big]$. Then, the longitudinal flux distribution $B(X,Y) \, {\hat Z}$ is also expressed with ${\bf K}_\perp$,
\begin{equation}
B(X,Y) =\sum_{{\bf K}_\perp} \, F_{{\bf K}_\perp} \exp\Big[i \, {\bf K}_\perp \cdot \,(X,Y)\Big].
\end{equation}
The Fourier component $|F_{{\bf K}_\perp}|$ with the smallest $|{\bf K}_\perp|\neq 0$ corresponds to VLFF which is, in the neutron scattering experiment, obtained from the integrated intensity of a diffraction peak \cite{VLFF}. Since VLFF measures the inhomogeneous part of the magnetic flux, the problem results in obtaining the spatially varying internal magnetic field.

\begin{figure}[t]
\begin{center}
\includegraphics[scale=0.35]{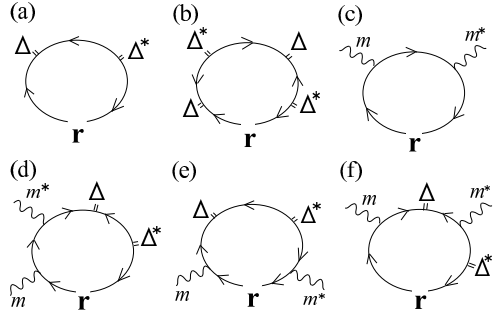}
\caption{Feynman diagrams for the SC and AFM contributions to the internal magnetic field. Here, the diagrams (a), (b), (c), and (d)-(f) denote the contributions proportional to $|\Delta|^2$, $|\Delta|^4$, $|m|^2$, and $|\Delta|^2|m|^2$, respectively. For the $|\Delta|^2|m|^2$ terms, not all the diagrams but only typical ones are shown: a diagram with two wavy lines on a single solid line (d), one with two wavy lines separated by the vertex (e), and one with two wavy lines separated by the pair field $|\Delta|$ (f). The two diagrams (d) and (e), and  the diagram (f) denote the internal fields arising from the free energies described by the diagram (b) in Fig. \ref{fig:Feynman_AF} and the diagram (c) in Fig. \ref{fig:Feynman_AF}, respectively.\label{fig:Feynman_current}}
\end{center}
\end{figure}

We derive a Maxwell equation relating the internal magnetic field ${\bf b}({\bf r})\equiv \nabla \times {\bf a}({\bf r})$ in Eq. (\ref{eq:flux}) to screening currents from the saddle point equation of the total free energy with respect to ${\bf A}$ $\delta F/\delta {\bf A}=0$, i.e., $ \big\langle  \delta{\cal H}_0/\delta {\bf A} \big\rangle\big|_{A=A_0} + \nabla \times (\nabla \times {\bf A}({\bf r}))/(4\pi) = 0$. Then, the Maxwell equation is given by
\begin{equation}\label{eq:Maxwell_fin}
-\frac{1}{4\pi}\nabla \times {\bf B}({\bf r})\simeq T \sum_{\varepsilon_n, \sigma} \sum_{\bf p} \big(- |e| {\bf v}_{\bf p} \big) G_{\varepsilon_n, \sigma}({\bf p};{\bf r}) -\nabla\times \Big(T \sum_{\varepsilon_n, \sigma} \sum_{\bf p} \, \mu_{\rm B}g\, \sigma \, G_{\varepsilon_n, \sigma}({\bf p};{\bf r})\Big) \hat{Z} ,
\end{equation}
where higher order terms in ${\bf A}$ have been dropped in Eq. (\ref{eq:Maxwell_fin}) because their contributions are negligibly small within the quasi-classical approximation, and $G_{\varepsilon_n, \sigma}({\bf p};{\bf r})$ is a Fourier transformation of the quasiparticle Green's function $G_{\varepsilon_n, \sigma}({\bf r},{\bf r'})$ and will be given later. In the right-hand side of Eq. (\ref{eq:Maxwell_fin}), the first term is the usual term expressing the current density, while the second term arises from the Zeeman term \cite{Houzet}. In this paper, we will use the perturbative expansion for the quasiparticle Green's function with respect to both $|\Delta|$ and $|m|$. Then, $G_{\varepsilon_n, \sigma}({\bf p};{\bf r})$ can be written as
\begin{equation}                                                                                                                                                                                                                                                                   
G_{\varepsilon_n, \sigma}({\bf p};{\bf r})=G^{(2)}_{\varepsilon_n, \sigma}({\bf p};{\bf r}) + G^{(4)}_{\varepsilon_n, \sigma}({\bf p};{\bf r})+ G^{(0,2)}_{\varepsilon_n, \sigma}({\bf p};{\bf r}) + G^{(2,2)}_{\varepsilon_n, \sigma}({\bf p};{\bf r})
\end{equation}
with
\begin{eqnarray}                                                                                                                                                                                                                                                                                              
G^{(2)}_{\varepsilon_n, \sigma}({\bf p};{\bf r})&=&- {\cal G}_{\varepsilon_n, \sigma}(-{\bf p}+\Pi_1) \, {\cal G}_{-\varepsilon_n, -\sigma}({\bf p}) \, {\cal G}_{\varepsilon_n, \sigma}(-{\bf p}+\Pi^\dagger_2)\, |w_{\bf p}|^2 \, \Delta({\bf s}_1) \Delta^\ast({\bf s}_2) \big|_{{\bf s}_i \rightarrow {\bf r}},  \nonumber\\
G^{(4)}_{\varepsilon_n, \sigma}({\bf p};{\bf r})&=&{\cal G}_{\varepsilon_n, \sigma}(-{\bf p}+{\bf \Pi}_1-{\bf \Pi}^\dagger_2) \, {\cal G}_{-\varepsilon_n, -\sigma}({\bf p}+{\bf \Pi}^\dagger_2) \,{\cal G}_{\varepsilon_n, \sigma}(-{\bf p}) \, {\cal G}_{-\varepsilon_n, -\sigma}({\bf p}+{\bf \Pi}_3) \, {\cal G}_{\varepsilon_n, \sigma}(-{\bf p}-{\bf \Pi}_3+{\bf \Pi}^\dagger_4) \nonumber\\
&&\times |w_{\bf p}|^4 \, \Delta ({\bf s}_1)\Delta^\ast ({\bf s}_2) \Delta({\bf s}_3) \Delta^\ast({\bf s}_4) \, \big|_{{\bf s}_i \rightarrow {\bf r}},\nonumber\\
G^{(0,2)}_{\varepsilon_n, \sigma}({\bf p};{\bf r})&=&\sum_{{\bf q},{\bf q}'} m({\bf q}) \, m^\ast({\bf q}') \, e^{i ({\bf q}-{\bf q}')\cdot{\bf r}} \, {\cal G}_{\varepsilon_n, \sigma}(-{\bf p}-{\bf q}') \,{\cal G}_{\varepsilon_n, {\overline \sigma}}(-{\bf p}-{\bf Q}_0) \, {\cal G}_{\varepsilon_n, \sigma}(-{\bf p}-{\bf q}),\nonumber\\
G^{(2,2)}_{\varepsilon_n, \sigma}({\bf p};{\bf r})&=&-\sum_{{\bf q},{\bf q}'} m({\bf q}) \, m^\ast({\bf q}') \, e^{i ({\bf q}-{\bf q}')\cdot{\bf r}} \, {\cal G}_{\varepsilon_n, \sigma}(-{\bf p}+{\bf \Pi}_1-{\bf q}') \, {\cal G}_{\varepsilon_n, \sigma}(-{\bf p}+{\bf \Pi}^\dagger_2-{\bf q})\nonumber\\
&\times& \bigg[ |w_{\bf p}|^2 \Big\{ {\cal G}_{\varepsilon_n, {\overline \sigma}}(-{\bf p}-{\bf Q}_0+{\bf \Pi}_1)  \, {\cal G}_{\varepsilon_n, \sigma}(-{\bf p}+{\bf \Pi}_1-{\bf q})  \, {\cal G}_{-\varepsilon_n, -\sigma}({\bf p}+{\bf q}) + ({\bf \Pi}_1 \leftrightarrow {\bf \Pi}^\dagger_2, \, {\bf q} \leftrightarrow {\bf q}') \Big\} \nonumber\\
&& + |w_{\bf p}|^2 \Big\{ {\cal G}_{-\varepsilon_n, -\sigma}({\bf p}+{\bf q}') \,{\cal G}_{-\varepsilon_n, -{\overline \sigma}}({\bf p}+{\bf Q}_0) \, {\cal G}_{-\varepsilon_n, -\sigma}({\bf p}+{\bf q}) \nonumber \\
&& \qquad + {\cal G}_{\varepsilon_n, {\overline \sigma}}(-{\bf p}+{\bf Q}_0+{\bf \Pi}_1) \, {\cal G}_{-\varepsilon_n, -{\overline \sigma}}({\bf p}-{\bf Q}_0) \, {\cal G}_{\varepsilon_n, {\overline \sigma}}(-{\bf p}+{\bf Q}_0+{\bf \Pi}^\dagger_2) \Big\} \nonumber\\
&& - w_{\bf p} w_{{\bf p}+{\bf Q}_0}\Big\{ {\cal G}_{\varepsilon_n, {\overline \sigma}}(-{\bf p}+{\bf Q}_0+{\bf \Pi}_1) \, {\cal G}_{-\varepsilon_n, -{\overline \sigma}}({\bf p}-{\bf Q}_0) \, {\cal G}_{-\varepsilon_n, -\sigma}({\bf p}-{\bf q}')  + ({\bf \Pi}_1 \leftrightarrow {\bf \Pi}^\dagger_2, \, {\bf q} \leftrightarrow {\bf q}') \Big\} \bigg] \nonumber\\
&& \times \Delta ({\bf s}_1) \, \Delta^\ast({\bf s}_2) \big|_{{\bf s}_i \rightarrow {\bf r}} ,
\end{eqnarray}  
where $G^{(n)}_{\varepsilon_n, \sigma}$ and $G^{(n,2)}_{\varepsilon_n, \sigma}$ denote the contributions proportional to $|\Delta|^n$ and $|\Delta|^n  |m|^2$, respectively. The contribution $G^{(0)}_{\varepsilon_n, \sigma}$ has been dropped since it only gives a spatially uniform flux. We note that the SC fourth order term $G^{(4)}$ is incorporated since, in the case with strong PPB of our interest, $|\Delta|$ is finite even on the $H_{c2}(T)$ curve at low temperatures due to the first order $H_{c2}$ transition and higher order SC contributions are not safely negligible. The contributions $G^{(2)}$, $G^{(4)}$, and $G^{(0,2)}$ are described by the Feynman diagrams (a), (b), and (c) in Fig. \ref{fig:Feynman_current}, respectively. The SC-AFM coupling term $G^{(2,2)}$ consists of the contributions described by the three types of Feynman diagrams (d)-(f) in Fig. \ref{fig:Feynman_current}. 
                                                                                                                                                                                        
Equation (\ref{eq:Maxwell_fin}) indicates that the magnetic flux can be written as
\begin{equation}\label{eq:internalfield}
{\bf B}({\bf r})=H{\hat Z}+ {\bf b}^{(2)}_\Delta({\bf r}) + {\bf b}^{(4)}_\Delta({\bf r}) + {\bf b}^{(0,2)}_m({\bf r}) + {\bf b}^{(2,2)}_{\Delta,m}({\bf r}),
\end{equation}                                                                                   
where ${\bf b}^{(l)}_\Delta({\bf r})$ and ${\bf b}^{(l,m)}_{\Delta,m}({\bf r})$ are the internal magnetic fields coming from $G^{(l)}_{\varepsilon_n,\sigma}$ and $G^{(l,m)}_{\varepsilon_n,\sigma}$, respectively.
First, the second order contribution ${\bf b}^{(2)}_\Delta({\bf r})$ will be calculated. By using the transformation (\ref{eq:parameter_int}), we have
\begin{eqnarray}\label{eq:magSC2nd}
&&T\sum_{\varepsilon_n} \int d\varepsilon({\bf p}) \, G^{(2)}_{\varepsilon_n, \sigma}({\bf p};{\bf r})= T\sum_{\varepsilon_n}  \,  |w_{\bf p}|^2  \frac{-2\pi i\, s_{\varepsilon_n}}{d_{\sigma, \Pi_1}^{(+)}  d_{\sigma, \Pi^\dagger_2}^{(+)}}  \Delta ({\bf s}_1) \Delta^\ast ({\bf s}_2) \big|_{{\bf s}_i \rightarrow {\bf r}} \nonumber\\
&&=\frac{1}{T_c} \int_0^\infty d\rho_1 \, d\rho_2 \frac{2\pi t}{\sinh\big[2\pi t (\rho_1+\rho_2)\big]} \frac{1}{2}\sum_{s_{\varepsilon_n} = \pm 1} (i s_{\varepsilon_n}) \, \exp\Big[i s_{\varepsilon_n} \overline{I} \, 2\sigma (\rho_1+\rho_2)\Big]  \nonumber\\
&& \times  |w_{\bf p}|^2 \, \exp\Big[i s_{\varepsilon_n} T_c^{-1}(\rho_1 {\bf v}_{\bf p}\cdot {\bf \Pi}_1 + \rho_2 {\bf v}_{\bf p}\cdot {\bf \Pi}^\dagger_2 )\Big] \, \Delta ({\bf s}_1) \Delta^\ast({\bf s}_2) \big|_{{\bf s}_i \rightarrow {\bf r}}.
\end{eqnarray}                                                                                                                                                                                                                                                                            

Further, with the help of Eq. (\ref{eq:formula2nd}), we obtain
\begin{eqnarray}\label{eq:flux2nd}
-\frac{{\bf b}^{(2)}_\Delta({\bf r})}{4\pi} &=& {\hat Z}  \frac{2|e|v_F r_H \, N(0)|\Delta|^2}{T_c} \int_0^\infty d\rho_1 \, d\rho_2 \, \sum_{{\bf K}_\perp \neq 0} T^{(2)}_{{\bf K}_\perp} \,e^{i{\bf K}_\perp\cdot (X,Y)} \Big\langle |w_{\bf p}|^2 \, \hat{b}^{(2)}_{{\bf K}_\perp,{\bf v}_{\bf p}}(\rho_{1,2}) \, e^{-\frac{|\eta|^2}{2} (\rho_1+\rho_2)^2}  \Big\rangle_{\rm FS} ,\nonumber\\
\hat{b}^{(2)}_{{\bf K}_\perp,{\bf v}_{\bf p}}(\rho_{1,2}) &=& \cos\Big(\frac{{\mbox {\boldmath $\eta$}} \cdot {\bf K}_\perp r_H }{\sqrt 2}(\rho_1-\rho_2)\Big) \bigg[ \, f_{\cos} \big(\rho_1+\rho_2,\rho_1+\rho_2 \big) \frac{\big({\bf v}_{\bf p} \times {\tilde {\bf K}} \, r_H \big)_{\hat{Z}}}{v_F|{\tilde {\bf K}}\, r_H|^2}  \sinh\Big(\frac{({\mbox {\boldmath $\eta$}}\times {\bf K}_\perp r_H )_z}{\sqrt 2}(\rho_1+\rho_2)\Big) \nonumber\\
&& + \frac{\overline{I}}{\pi}\frac{r_H}{\xi_0} f_{\sin} \big(\rho_1+\rho_2,\rho_1+\rho_2 \big)  \cosh\Big(\frac{({\mbox {\boldmath $\eta$}}\times {\bf K}_\perp r_H )_z}{\sqrt 2}(\rho_1+\rho_2)\Big) \bigg],
\end{eqnarray}
where $v_F$ is the Fermi velocity in the pure two dimensional system and $\xi_0$ is the SC coherence length at $T=0$ defined by $\xi_0=v_F/(2\pi \, T_c)$. Here, the term proportional to $(i|e|{\bf v}_{\bf p})$ is transformed, with the rotation operator $\nabla \times \cdots$, in the form
\begin{eqnarray}\label{eq:v-rotation}                                               
\sum_{{\bf K}_\perp \neq 0} \, i{\bf v}_{\bf p} \, f_{{\bf K}_\perp}({\bf v}_{\bf p})e^{i{\bf K}_\perp \cdot (X,Y)} 
&\simeq& \nabla \times \bigg( \sum_{{\bf K}_\perp \neq 0}\frac{({\bf v}_{\bf p}\times {\tilde {\bf K}})_{\hat Z}}{|{\tilde {\bf K}}|^2} \, f_{{\bf K}_\perp}({\bf v}_{\bf p})e^{i{\bf K}_\perp \cdot (X,Y)} \bigg)\hat{Z},
\end{eqnarray} 
where ${\tilde {\bf K}}$ is defined by ${\tilde {\bf K}}=(K_x,K_y,0)$ for ${\hat Z}\parallel {\hat z}$ and ${\tilde {\bf K}}=(\gamma^{-1/2}K_y,0,\gamma^{1/2}K_x)$ for ${\hat Z}\parallel {\hat y}$.

To check the result obtained here, we consider the internal field near $T_c$ at nearly zero magnetic field, i.e., in the GL region. We take the limit $t\rightarrow 1$ and $H \rightarrow 0$ (or equivalently, $\eta \rightarrow 0$) in Eq. (\ref{eq:flux2nd}) in ${\bf H}\parallel c$ case. Then, the contribution arising from the Zeeman term, which is proportional to $\overline{I}$ in Eq. (\ref{eq:flux2nd}), vanishes, and the remnant orbital contribution is expressed as
\begin{eqnarray}\label{eq:flux2nd_GL}                                                           
\lim_{T\rightarrow T_c, \, H\rightarrow 0} \frac{-{\bf b}^{(2)}_\Delta({\bf r})}{4\pi}
&=&\frac{2\pi|e|v_F^2N(0)|\Delta|^2}{T_c^2}  \sum_{{\bf K}_\perp} T^{(2)}_{{\bf K}_\perp} \,e^{i{\bf K}_\perp\cdot {\bf r}_\perp} \int_0^\infty d\rho \frac{\rho_1+\rho_2}{\sinh \big[2\pi \rho_1+\rho_2 \big]} \bigg\langle |w_{\bf p}|^2 \frac{({\bf v}_{\bf p}\times {\bf K}_\perp)_z^2}{v_F^2|{\bf K}_\perp|^2} \bigg\rangle_{\rm FS} \nonumber\\
&=&C_{\rm GL}\frac{|\Delta|^2}{T_c^2} \, \sum_{{\bf K}_\perp} T^{(2)}_{{\bf K}_\perp} \,e^{i{\bf K}_\perp\cdot {\bf r}_\perp} \bigg\langle |w_{\bf p}|^2 \frac{({\bf v}_{\bf p}\times {\bf K}_\perp)_z^2}{v_F^2 |{\bf K}_\perp|^2} \bigg\rangle_{\rm FS}
\end{eqnarray}   
with 
\begin{equation}
C_{\rm GL}\equiv\frac{4|e|v_{\rm F}^2 N(0)}{(2\pi )^2} \frac{7}{8}\zeta(3),
\end{equation}
where $\int_0^\infty d\rho \, \rho^2/\sinh[2\pi \, \rho] = 4\sum_{n \ge 0} 1/[2\pi(2n+1)]^3$ is used. Equation (\ref{eq:flux2nd_GL}) is a well-known familiar form of the internal magnetic field obtained in the GL region \cite{Werthamer}. 
      
Next, we will calculate the fourth order contribution ${\bf b}^{(4)}_\Delta({\bf r})$. By carrying out the integral $\int d\varepsilon({\bf p})$ and using the transformation (\ref{eq:parameter_int}), we have                                                                   
\begin{eqnarray}\label{eq:flux4th_0}
&&T\sum_{\varepsilon_n} \int d\varepsilon({\bf p}) \, G^{(4)}_{\varepsilon_n, \sigma}({\bf p};{\bf r}) = T\sum_{\varepsilon_n}  \, (-2\pi  \, i s_{\varepsilon_n}) |w_{\bf p}|^4 \bigg[ \frac{1}{d_{\sigma,\Pi_1}^{(+)} \,  d_{\sigma,\Pi_2^\dagger}^{(+)} \,  d_{\sigma,\Pi_3}^{(+)} \,  d_{\sigma,\Pi_4^\dagger}^{(+)}}  \nonumber\\
&& + \frac{1}{d_{\sigma,\Pi_1}^{(+)} \,  d_{\sigma,\Pi_2^\dagger}^{(+)} \, d_{\sigma,\Pi_4^\dagger}^{(+)} \, d_{\sigma,\Pi_2^\dagger-\Pi_3+\Pi_4^\dagger}^{(+)} }+ \frac{1}{d_{\sigma,\Pi_1}^{(+)} \,  d_{\sigma,\Pi_3}^{(+)} \, d_{\sigma,\Pi_4^\dagger}^{(+)} \, d_{\sigma,\Pi_1-\Pi_2^\dagger+\Pi_3}^{(+)}} \bigg] \Delta({\bf s}_1) \Delta^\ast ({\bf s}_2) \, \Delta ({\bf s}_3) \Delta^\ast ({\bf s}_4) \big|_{{\bf s}_i \rightarrow {\bf r} }\nonumber\\
&=&-\frac{1}{T_c^3} \frac{1}{2}\sum_{s_{\varepsilon_n}=\pm 1} (i s_{\varepsilon_n}) \, \prod _{i=1}^4 \int_0^\infty d\rho_i \frac{2\pi t}{\sinh \big[2\pi t (\sum_{i=1}^4 \rho_i) \big]} \exp \Big[ i s_{\varepsilon_n} \overline{I} 2\sigma (\sum_{i=1}^4 \rho_i)\Big] \nonumber\\
&&\times \sum_{j=1}^3 \, |w_{\bf p}|^4 \exp\bigg[ i \, s_{\varepsilon_n} T_c^{-1}\Big({\bf v}_{\bf p}\cdot{\bf \Pi}_1 \, \alpha_j + {\bf v}_{\bf p}\cdot{\bf \Pi}^\dagger_2 \, \beta_j + {\bf v}_{\bf p}\cdot{\bf \Pi}_3 \, \gamma_j + {\bf v}_{\bf p}\cdot{\bf \Pi}^\dagger_4 \, \delta_j \Big)\bigg] \Delta ({\bf s}_1) \Delta^\ast ({\bf s}_2) \, \Delta ({\bf s}_3) \Delta^\ast({\bf s}_4) \big|_{{\bf s}_i \rightarrow {\bf r}},
\end{eqnarray}                                                                                                                                                                                                                                         
where the coefficients $\alpha_j, \, \beta_j, \, \gamma_j$, and $\delta_j$ are shown in table \ref{table:flux4th}. With the combined use of Eqs. (\ref{eq:formula4th}) and (\ref{eq:v-rotation}), we obtain
\begin{eqnarray}\label{eq:flux4th}
-\frac{{\bf b}^{(4)}_\Delta({\bf r})}{4\pi \, C_{\rm GL}} &=-& {\hat Z} \, \frac{8\pi}{7 \zeta (3)} \frac{ r_H }{\xi_0} \frac{|\Delta|^4 }{T_c^4}\, \frac{k} {\sqrt{ 2\pi}}\sum_{{\bf K}_\perp \neq 0} T^{(4)}_{{\bf K}_\perp} \, \,e^{i{\bf K}_\perp \cdot (X,Y)} \, \sum_{l_1,l_2=-\infty}^\infty T^{(4)}(l_1,l_2,{\bf K}_\perp) \prod _{i=1}^4 \int_0^\infty d\rho_i \nonumber\\
&&\times \sum_{j=1}^3 \bigg\langle |w_{\bf p}|^4 \, \hat{b}^{(4)}_{{\bf k}_\perp,{\bf p}}(\rho_j) \, {\rm exp}\bigg[ -\frac{|\eta|^2}{2} \Big( \alpha_j ^2 +\beta_j ^2 +\gamma_j ^2 +\delta_j ^2 +(\alpha_j+\gamma_j)(\beta_j+\delta_j) \Big) \bigg] \nonumber\\
&& \times \, {\rm exp}\bigg[ -\frac{1}{8} {\rm Re}\big( \eta^2 \big) \Big( (\alpha_j -\gamma_j )^2  + (\beta_j -\delta_j )^2  \Big) \bigg] \, {\rm exp}\bigg[ i \, \frac{1}{8} {\rm Im}\big( \eta^2 \big) \Big( (\alpha_j -\gamma_j )^2  - (\beta_j -\delta_j )^2  \Big) \bigg]\bigg\rangle_{\rm FS}, \nonumber\\
\hat{b}^{(4)}_{{\bf k}_\perp,{\bf p}}(\rho_j) &=&f_{\cos} \big(\sum_{i=1}^4 \rho_i,\sum_{i=1}^4 \rho_i \big) \frac{({\bf v}_{\bf p}\times {\tilde{\bf K} r_H})_{\hat Z}}{v_F|{\tilde{\bf K}} r_H|^2} \Big[ \cos\big(V_j \big) \sinh\big( W_j \big) -i\, \sin\big(V_j \big) \cosh\big( W_j \big)\Big]\nonumber\\
&&+ \frac{\overline{I}}{\pi}\frac{r_H}{\xi_0} f_{\sin} \big(\sum_{i=1}^4 \rho_i,\sum_{i=1}^4 \rho_i \big) \Big[\cos\big(V_j \big) \cosh\big( W_j \big) -i\, \sin\big( V_j \big) \sinh\big( W_j \big)\Big]                                        
\end{eqnarray} 
with
\begin{eqnarray}
V_j&=&\frac{({\mbox {\boldmath $\eta$}}\cdot {\bf K}_\perp r_H)}{2{\sqrt 2}}\big\{(\alpha_j+\gamma_j)-(\beta_j+\delta_j) \big\} - {\rm Im}\big(\eta\big)\frac{k}{\sqrt 2} \big\{(l_1-l_2)(\alpha_j-\gamma_j) - (l_1+l_2+\frac{K_y r_H}{k})(\beta_j-\delta_j) \big\}, \nonumber\\
W_j&=&\frac{({\mbox {\boldmath $\eta$}}\times {\bf K}_\perp r_H)_z}{2{\sqrt 2}}\big\{(\alpha_j+\gamma_j)+(\beta_j+\delta_j) \big\} - {\rm Re}\big(\eta\big)\frac{k}{\sqrt 2} \big\{(l_1-l_2)(\alpha_j-\gamma_j) + (l_1+l_2+\frac{K_y r_H}{k})(\beta_j-\delta_j) \big\} .
\end{eqnarray}

\begin{table}[t]
\caption{ Coefficients $\alpha_j, \, \beta_j, \, \gamma_j, \, \delta_j$ in Eqs. (\ref{eq:flux4th_0}) and (\ref{eq:flux4th})\label{table:flux4th}}
\begin{tabular}{>{$}c<{$}>{$}c<{$}>{$}c<{$}>{$}c<{$}>{$}c<{$}}
\hline
\hline
j & \alpha_j & \beta_j & \gamma_j & \delta_j  \\
\hline
1 & \rho_1         & \rho_2        &  \rho_3         & \rho_4 \\ 
2 & \rho_1         & \rho_2+\rho_3 &  -\rho_3        & \rho_3+ \rho_4 \\ 
3 & \rho_1+\rho_2  & -\rho_2       &  \rho_2+\rho_3  & \rho_4 \\ 
\hline
\hline
\end{tabular}
\end{table}

Concerning the internal magnetic field induced by the AFM order, the leading order $|m|^2$ term which does not include $|\Delta|$ is straightforwardly calculated as 
\begin{eqnarray}\label{eq:flux02nd}
-\frac{{\bf b}^{(0,2)}_m({\bf r})}{4\pi C_{\rm GL}} &=& {\hat Z}\frac{8\pi}{7 \zeta (3)} \frac{ r_H^2 }{\xi_0^2}\frac{\overline{I}}{\pi} \int_0^\infty d\rho_1 \, d\rho_2 \, f_{\sin} (\rho_1+\rho_2,0) \cos\Big( \delta_{\rm IC} (\rho_1+\rho_2) \Big)  \nonumber\\
&& \times  \frac{1}{T_c^2} \sum_{{\bf q},{\bf q}'} m({\bf q}) \, m^\ast({\bf q}') \, e^{i ({\bf q}-{\bf q}')\cdot{\bf r}} \bigg\langle \cos \Big( T_c^{-1}{\bf v}_{\bf p}\cdot {\bf q} \, \rho_1 + T_c^{-1}{\bf v}_{\bf p}\cdot {\bf q}' \, \rho_2 \Big) \bigg\rangle_{\rm FS}.
\end{eqnarray}   
We note that, when the AFM moment is perpendicular to a magnetic field (${\overline \sigma}=-\sigma$), ${\bf b}^{(0,2)}_m({\bf r})$ vanishes since $f_{\sin}(x,0)=0$ in ${\bf m}\perp {\bf H}$. 
                                                                              
The internal magnetic field brought by the SC-AFM coupling ${\bf b}^{(2,2)}_{\Delta,m}({\bf r})$ can be derived in the same manner as that used in obtaining ${\bf b}^{(2)}_\Delta ({\bf r})$. $T\sum_{\varepsilon_n} \int d\varepsilon({\bf p}) \,  G^{(2,2)}_{\varepsilon_n, \sigma}({\bf p};{\bf r})$ is calculated as
\begin{eqnarray}\label{eq:flux2nd2nd_0}                                                                                                                                                                                
&&T\sum_{\varepsilon_n} \int d\varepsilon({\bf p}) \,  G^{(2,2)}_{\varepsilon_n, \sigma}({\bf p};{\bf r}) = -2\pi \, T\sum_{\varepsilon_n} (i \, s_{\varepsilon_n})  \sum_{{\bf q},{\bf q}'} m({\bf q}) \, m^\ast({\bf q}') \, e^{i ({\bf q}-{\bf q}')\cdot{\bf r}} \bigg( |w_{\bf p}|^2 \, \bigg[ \frac{1}{d_{\sigma,\Pi_1}^{(+)} d_{\sigma,\Pi_2^\dagger}^{(+)} \, d_{\sigma,\Pi_1+q-q'}^{(+)} \, d_{+,q'}^{(-)}} \nonumber\\
&& + \frac{1}{d_{\sigma,\Pi_1}^{(+)} d_{\sigma,\Pi_2^\dagger}^{(+)} d_{+,q'}^{(-)} d_{+,q}^{(-)}} + \frac{1}{d_{\sigma,\Pi_2^\dagger}^{(+)} d_{+,q'}^{(-)} d_{+,q}^{(-)} d_{+,\Pi_1-\Pi_2^\dagger+q}^{(-)}} + ({\bf \Pi}_1 \leftrightarrow {\bf \Pi}^\dagger_2, \, {\bf q} \leftrightarrow {\bf q}') \bigg] + |w_{\bf p}|^2 \, \bigg[ \frac{1}{d_{\sigma,\Pi_1}^{(+)} d_{\sigma,\Pi_2^\dagger}^{(+)} \, d_{\sigma,\Pi_1+q-q'}^{(+)} \, d_{+,q}^{(+)}} \nonumber\\
&& + \frac{1}{d_{\sigma,\Pi_1}^{(+)} d_{\sigma,\Pi_2^\dagger}^{(+)} \, d_{\sigma,\Pi_2^\dagger-q+q'}^{(+)} \, d_{+,q'}^{(+)}} + \frac{1}{d_{\sigma,\Pi_1}^{(+)} d_{\sigma,\Pi_2^\dagger}^{(+)} d_{+,q'}^{(+)} d_{+,q}^{(+)}}    + \frac{1}{d_{\overline{\sigma},\Pi_1}^{(-)} d_{\overline{\sigma},\Pi_2^\dagger}^{(-)} d_{+,q'}^{(-)} d_{+,q}^{(-)}} + \frac{1}{d_{\overline{\sigma},\Pi_1}^{(-)} d_{+,q'}^{(-)} d_{+,q}^{(-)} d_{+,\Pi_1-\Pi_2^\dagger+q}^{(-)}} \nonumber\\
&& + \frac{1}{d_{\overline{\sigma},\Pi_2^\dagger}^{(-)} d_{+,q'}^{(-)} d_{+,q}^{(-)} d_{+,-\Pi_1+\Pi_2^\dagger+q'}^{(-)}} \bigg] - w_{\bf p}w_{{\bf p}+{\bf Q}_0} \, \bigg[ -\frac{1}{d_{\overline{\sigma},\Pi_1}^{(-)} d_{+,q'}^{(-)} d_{+,q}^{(+)} d_{+,\Pi_1-\Pi_2^\dagger+q}^{(-)}} - \frac{1}{d_{\sigma,\Pi_2^\dagger}^{(+)} d_{+,q'}^{(-)} d_{+,q}^{(+)} d_{+,\Pi_1-\Pi_2^\dagger+q}^{(-)}} \nonumber\\
&& - \frac{1}{d_{\sigma,\Pi_2^\dagger}^{(+)} \, d_{\sigma,\Pi_1+q-q'}^{(+)} \, d_{+,q'}^{(-)} d_{+,q}^{(+)}} + ({\bf \Pi}_1 \leftrightarrow {\bf \Pi}^\dagger_2, \, {\bf q} \leftrightarrow {\bf q}') \bigg] \bigg) \, \Delta ({\bf s}_1) \, \Delta^\ast({\bf s}_2) \big|_{{\bf s}_i\rightarrow {\bf r}} \nonumber\\
&=&-\frac{1}{T_c^3} \frac{1}{2}\sum_{s_{\varepsilon_n}=\pm 1} (i s_{\varepsilon_n}) \, \prod _{i=1}^4 \int_0^\infty d\rho_i \frac{2\pi t}{\sinh \big[2\pi t \,(\sum_{i=1}^4 \rho_i) \big]} \times \sum_{j=1}^{12}\exp \Big[ i s_{\varepsilon_n} \overline{I} \Big( 2\sigma e_j + (\sigma+{\overline \sigma}) (\sum_{i=1}^4 \rho_i - |e_j|)\Big)\Big] \nonumber\\
&\times& \exp \Big[ - i s_{\varepsilon_n} \delta_{\rm IC} (c_j+d_j )\Big] \exp\bigg[ i \, s_{\varepsilon_n} T_c^{-1} \Big({\bf v}_{\bf p}\cdot{\bf \Pi}_1 \, a_j + {\bf v}_{\bf p}\cdot{\bf \Pi}^\dagger_2 \, b_j - {\bf v}_{\bf p}\cdot{\bf q} d_j - {\bf v}_{\bf p}\cdot{\bf q}' c_j \Big)\bigg] \nonumber\\
&&\times  \bigg( |w_{\bf p}|^2 \, \Big\{ \delta_{j,1-3}\Big[1+({\bf \Pi}_1 \leftrightarrow {\bf \Pi}^\dagger_2, \, {\bf q} \leftrightarrow {\bf q}')\Big]+\delta_{j,4-9} \Big\} +w_{\bf p} w_{{\bf p}+{\bf Q}_0} \, \delta_{j,10-12}\Big[1+({\bf \Pi}_1 \leftrightarrow {\bf \Pi}^\dagger_2, \, {\bf q} \leftrightarrow {\bf q}')\Big] \bigg) \Delta ({\bf s}_1) \, \Delta^\ast({\bf s}_2) \big|_{{\bf s}_i \rightarrow {\bf r}},
\end{eqnarray}                                                                                                                                                                                                                                                                                                                
where $ e_j =a_j + b_j$ and the coefficients  $p_j, \,  a_j, \, b_j, \,  c_j$, and  $d_j $ are shown in table \ref{table:flux2nd2nd}. Then, we obtain

\begin{eqnarray}\label{eq:flux2nd2nd}
-\frac{{\bf b}^{(2,2)}_{\Delta,m}({\bf r})}{4\pi C_{\rm GL}} &=-& \frac{8\pi}{7 \zeta (3)} \frac{r_H }{\xi_0} \frac{|\Delta|^2}{T_c^4}  \, \sum_{{\bf q},{\bf q}'} m({\bf q}) \, m^\ast({\bf q}') \, \sum_{{\bf K}_\perp \neq 0} T^{(2)}_{{\bf K}_\perp } \, \,e^{i{\bf K}_\perp \cdot (X,Y)} \, e^{i ({\bf q}-{\bf q}')\cdot{\bf r}} \nonumber\\
&\times& \prod _{i=1}^4 \int_0^\infty d\rho_i  \sum_{j=1}^{12}  \,p_j  \cos\Big( \delta_{\rm IC} (c_j+d_j) \Big) \Big\langle |w_{\bf p}|^2 \, \hat{b}^{(2,2)}_{{\bf K}_\perp,{\bf v}_{\bf p}}(\rho_j;{\bf q},{\bf q'}) \, e^{ -\frac{|\eta|^2}{2} (a_j+b_j)^2 } \Big\rangle_{\rm FS}, \nonumber\\
\hat{b}^{(2,2)}_{{\bf K}_\perp ,{\bf v}_{\bf p}}(\rho_j;{\bf q},{\bf q'})&=& \cos\Big(\frac{({\mbox {\boldmath $\eta$}} \cdot {\bf K}_\perp r_H )}{{\sqrt 2}}(a_j-b_j)+\frac{{\bf v}_{\bf p}\cdot{\bf q'}}{T_c} c_j +\frac{{\bf v}_{\bf p}\cdot{\bf q}}{T_c} d_j \Big) \nonumber\\
&\times& \bigg[ f_{\cos} \big(\sum_{i=1}^4 \rho_i, \, e_j\big)  \frac{({\bf v}_{\bf p}\times {\tilde{\bf K'} r_H})_{\hat Z}}{v_F|{\tilde{\bf K}}'r_H|^2} \sinh\Big( \frac{({\mbox {\boldmath $\eta$}}\times {\bf K}_\perp r_H)_z}{{\sqrt 2}}(a_j+b_j) \Big) \nonumber\\
&& + \frac{\overline{I}}{\pi}\frac{r_H}{\xi_0} f_{\sin} \big(\sum_{i=1}^4 \rho_i, \, e_j \big) \cosh\Big( \frac{({\mbox {\boldmath $\eta$}}\times {\bf K}_\perp r_H)_z}{{\sqrt 2}}(a_j+b_j) \Big)\bigg] ,
\end{eqnarray} 
where the relation $w_{{\bf p}+{\bf Q}_0}=w_{\bf p}$ is used and ${\tilde{\bf K}}'$ is defined by ${\tilde{\bf K}}'=(K_x+q_x-q'_x,K_y+q_y-q'_y,0)$ for ${\hat Z}\parallel {\hat z}$ and ${\tilde{\bf K}}'=(\gamma^{-1/2}K_y+q_x-q'_x,0,\gamma^{1/2}K_x+q_z-q'_z)$ for ${\hat Z}\parallel {\hat y}$.
                                                                                                                                                                                                                                                                               
\begin{table}[t]
\caption{Coefficients $p_j, \,   a_j, \, b_j, \, c_j, \,  d_j$ in Eqs. (\ref{eq:flux2nd2nd_0}) and (\ref{eq:flux2nd2nd})\label{table:flux2nd2nd}}
\begin{tabular}{>{$}c<{$}>{$}c<{$}>{$}c<{$}>{$}c<{$}>{$}c<{$}>{$}c<{$}>{$}c<{$}}
\hline
\hline
j & p_j & a_j & b_j & c_j & d_j \\
\hline
1 & 2  &  -\rho_2        &  \rho_1+\rho_2  & \rho_4          &  \rho_2+\rho_3 \\ 
2 & 2  &  \rho_1+\rho_3  &  \rho_2         & \rho_3+\rho_4   &  -\rho_3       \\ 
3 & 2  &  \rho_1         &  \rho_2         & \rho_4          &  \rho_3        \\ 
4 & 1  &  \rho_1         &  \rho_2+\rho_3  &  -\rho_3-\rho_4 &  \rho_3        \\
5 & 1  &  \rho_1+\rho_3  &  \rho_2         &  \rho_3         &  -\rho_3-\rho_4\\
6 & 1  &  \rho_1         &  \rho_2         &  -\rho_3        &  -\rho_4       \\
7 & 1  &  -\rho_1-\rho_2 &  \rho_2         &  \rho_3         &  \rho_2+\rho_4 \\
8 & 1  &  \rho_2         &  -\rho_1-\rho_2 &  \rho_2+\rho_3  &  \rho_4        \\
9 & 1  &  -\rho_1        &  -\rho_2        &  \rho_3         &  \rho_4        \\
10& -2 &  -\rho_1-\rho_2 &  \rho_2         &  \rho_3         &  \rho_2-\rho_4 \\
11& -2 &  -\rho_2        &  \rho_1+\rho_2  &  \rho_3         &  \rho_2-\rho_4 \\
12& -2 &  \rho_1         &  \rho_2         &  \rho_1+\rho_3  &  -\rho_1-\rho_4\\
\hline
\hline
\end{tabular}
\end{table}

In this paper, we use the following form of the reciprocal lattice vector: 
\begin{equation}
{\bf K}_\perp r_H =m_1 \, (\pi/k,0) + m_2 \, (0,k)
\end{equation}
with integers $m_1$ and $m_2$. Then, the summation for ${\bf K}_\perp$ is replaced with $\sum_{m_1,m_2}$, and $T_{{\bf K}_\perp}^{(2)}$ in Eq. (\ref{eq:flux2nd}) and $T_{{\bf K}_\perp}^{(4)}$ in Eq. (\ref{eq:flux4th}) are concretely given as follows:
\begin{eqnarray}
T_{{\bf K}_\perp}^{(2)}&=&\delta_{m_1+m_2 , {\rm even}} \cos\Big(\frac{\pi}{2}m_2(m_1+m_2)\Big) \exp\Big[-\frac{1}{4}\Big(\frac{(m_1 \pi)^2}{k^2}+ (m_2 k)^2 \Big)\Big],\nonumber\\
T_{{\bf K}_\perp}^{(4)}&=&\delta_{m_1+m_2 , {\rm even}} \exp \Big[-i\frac{\pi}{4}m_2(m_1+m_2)\Big] \exp\Big[-\frac{1}{8}\Big(\frac{(m_1 \pi)^2}{k^2}+ (m_2 k)^2 \Big) \Big].
\end{eqnarray}                                                                                                 
Since the triangular lattice $k=\sqrt{\pi \sqrt{3}}$ is assumed, there are six $(m_1,m_2)$ combinations with the shortest $|{\bf K}_\perp|\neq 0$, $\pm(1,1), \, \pm(1,-1)$, and $\pm(2,0)$, so that the equality $|F_{\pm 1,1}|=|F_{1,\pm 1}|=|F_{\pm 2,0}|$ is satisfied in an isotropic system. However, in the present system, an asymmetry occurs in the six $|F_{m_1,m_2}|$ components because of the anisotropy originating from the $d$-wave pairing symmetry or the layered crystal structure. As a result, the equality $|F_{\pm 2,0}|=|F_{1,\pm 1}|$ is {\it not} satisfied any longer, while the equality $|F_{\pm 1,1}|=|F_{1,\pm 1}|$ is still satisfied. Since the four $(m_1,m_2)$ points with the same $|F_{m_1,m_2}|$ value, $\pm (1,1)$ and $\pm (1,-1)$, correspond to the spots with the strongest neutron scattering intensity in the experiments, we will calculate $|F_{1,1}|$ which is given by    
\begin{equation}\label{eq:formfactor}                                                                                                 
F_{1,1}=\frac{1}{V} \int_{\bf r} \Big[{\bf b}^{(2)}_\Delta({\bf r})+{\bf b}^{(4)}_\Delta({\bf r})+{\bf b}^{(0,2)}_m({\bf r})+{\bf b}^{(2,2)}_{\Delta,m}({\bf r})\Big] \exp \Big[ -i \, (\pi/k,k)\cdot (X,Y)\Big].
\end{equation}
The form factor $|F_{1,1}|$ is easily obtained by using Eqs. (\ref{eq:flux2nd}), (\ref{eq:flux4th}), (\ref{eq:flux02nd}), and (\ref{eq:flux2nd2nd}) since ${\bf b}^{(l)}_\Delta$ and ${\bf b}^{(l,m)}_{\Delta,m}$ in these equations are already expanded with the reciprocal lattice vector ${\bf K}_\perp$. In the results of our calculation below, $|F_{1,1}|$ is normalized by $4\pi C_{\rm GL}$.

In the present theory, the layered crystal structure is reflected in the anisotropy of the Fermi velocity vector ${\bf v}_{\bf p}$. Although the expression of ${\bf v}_{\bf p}$ should be derived from the realistic dispersion, for brevity we use the following simplified form of ${\bf v}_{\bf p}$: 
\begin{equation}\label{eq:velocity}
{\bf v}_{\bf p}=v_{\rm F} \big[1-(J/E_F)\big(1-\cos(p_z d)\big)\big]^{1/2}( {\hat x} \, \cos\phi+ {\hat y} \, \sin\phi )+ J d \sin(p_z d){\hat z},
\end{equation}
where $J$ is an interlayer coupling constant. The Fermi velocity vector introduced here corresponds to that of the corrugated cylindrical Fermi surface. Then, the anisotropy of the SC coherence length $\gamma$ can be expressed as $\gamma = 2 \sqrt{1-J/E_F}/(\pi J/E_F)$ and the angle average on the Fermi surface is defined by
\begin{equation}
\big\langle A \big\rangle_{\rm FS}  = \int_0^{2\pi}\frac{d \, \phi}{2\pi} \, \int_{-\pi}^{\pi}\frac{d \, (p_z d)}{2\pi}  \, A.
\end{equation}
In the numerical calculation below, $J/E_F=0.2$ and $p_F d=\pi$ are used, and then, $\gamma=2.85$ is obtained.

In the numerical calculation, the magnetic field $H$ is normalized by the orbital limiting field in the pure two dimensional system $H^{\rm (orb)}_{\rm 2D}(0)=0.28/(|e| \xi_0^2)$. Then, the two field-dependent energy scales $\mu_{\rm B}gH$ and $v_F /(2\pi \, r_H)$ which correspond to the paramagnetic pair-breaking effect and the orbital one, respectively are expressed as follows:
\begin{eqnarray}
\overline{I}=\frac{\mu_{\rm B}gH}{T_c}&=& h \, \alpha_{M,j} \, \frac{\pi}{2e^{\gamma_E}} \frac{H^{\rm (orb)}_{\rm 2D}(0)}{ H^{\rm (orb)}_j(0)}=0.882 \,\alpha_{M,j} \, \frac{H^{\rm (orb)}_{\rm 2D}(0)}{ H^{\rm (orb)}_j(0)} \, h ,\nonumber\\
\frac{\xi_0}{r_H}&=&\frac{v_F }{2\pi \, r_H \, T_c}= \xi_0 {\sqrt 2|e| H^{\rm (orb)}_{\rm 2D}(0)} {\sqrt h}= 0.748 \, \sqrt{h} ,
\end{eqnarray}  
where the Maki parameter $\alpha_{M,j}=\sqrt{2} \, H^{\rm (orb)}_j(0)/H_p(0)$ measures the strength of the paramagnetic pair-breaking effect and $h=H/H^{\rm (orb)}_{\rm 2D}(0)$ is the normalized magnetic field. Here, $H_p(0)=\pi T_c/(\sqrt{2}e^{\gamma_E}\mu_{\rm B}g)\simeq 1.2 \, T_c/\mu_{\rm B}g$ is the Pauli limiting field at $T=0$, where $\gamma_E=0.577$ is the Euler constant, while $H^{\rm (orb)}_j(0)$ is the orbital limiting field at $T=0$ for fields parallel to the j direction. For the parameters used in our calculation, $H^{\rm (orb)}_c(0)=1.266 \, H^{\rm (orb)}_{\rm 2D}(0) $ and  $H^{\rm (orb)}_{ab}(0)=4.871 \, H^{\rm (orb)}_{\rm 2D}(0) $ are obtained.

\section{Antiferromagnetic fluctuation in ${\bf H}\parallel c$}
In this section, we consider effects of the PPB-induced AFM fluctuation on VLFF in ${\bf H}\parallel c$ case and discuss the origin of the anomalous field dependence of VLFF observed in CeCoIn${}_5$ \cite{VLFF, White}. Throughout this section, $\alpha_{\rm M,c}=5.8$ is used in obtaining main results. The temperature-field phase diagram obtained by minimizing ${\cal F}_{\rm GL}(\Delta,m=0)$ is shown in the inset in Fig. \ref{fig:AFfluc}, where the $H_{c2}$ transition at low temperatures is of first order, and a possible FFLO vortex lattice state appearing just below the $H_{c2}(T)$ curve as a narrow HFLT region \cite{Kumagai_06,RI07c} is neglected because it does not affect the SC properties in the field range of our interest. In fact, we will demonstrate later in Sec. IV C that, even if the FFLO vortex lattice is taken into account, VLFF {\it in ${\bf H}\parallel c$ case} will be hardly affected by a FFLO spatial structure. Below, we will show results of our calculation in two cases, ${\bf m}\perp {\bf H}$ and ${\bf m}\parallel {\bf H}$, since the direction of ${\bf m}$ relative to the external magnetic field ${\bf H}$ has not been confirmed in ${\bf H} \parallel c$ case.

\subsection{Paramagnetic pair-breaking effect on VLFF}                                                                                                                                                                                                                                                                                                                                                           
\begin{figure}[t]                                                                                                                                                                                                                                                                                                                                                                                                                                                                                                                                                                                                                                                                                                                                                                                                                                                                                                                                                                                                                                                    
\begin{center}
\includegraphics[scale=0.5]{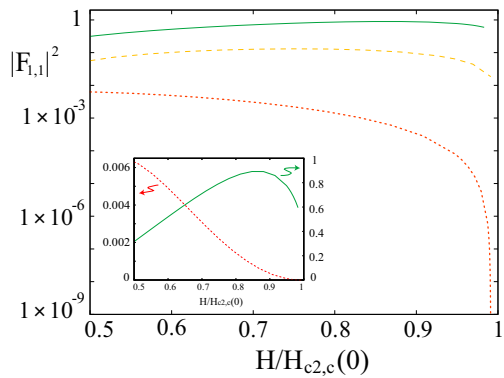}
\caption{(Color online) Logarithmic plots of VLFF $|F_{1,1}|^2$ at $t=0.1$ as a function of magnetic field scaled by the upper critical field at $T=0$ $H_{c2,c}(0)$ for various values of Maki parameter $\alpha_{ M,c}$. Here, $|F_{1,1}|^2$ is normalized by $(4\pi C_{\rm GL})^2$. Dotted, dashed, and solid curves correspond to the cases with $\alpha_{\rm M,c}=0.01$, $\alpha_{\rm M,c}=3.5$, and $\alpha_{\rm M,c}=5.8$, respectively. The form factor $|F_{1,1}|^2$ is enhanced by strong PPB at high fields, and as a result, the slope changes its sign from negative to positive. The inset shows the regular plot of the $|F_{1,1}|^2$ curves shown in the main panel, where the results for $\alpha_{M,c}=0.01$ (dotted curve) and $\alpha_{M,c}=5.8$ (solid one) are measured by the left and right vertical axes, respectively. \label{fig:vlff_pauli}}
\end{center}                                                                                                                                                                                                                                                                                                                                                                                                                                                                        
\end{figure}

\begin{figure}[t]
\begin{center}
\includegraphics[scale=0.65]{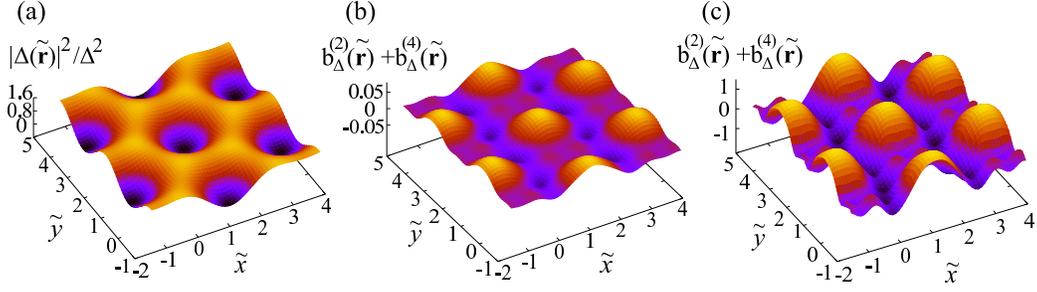}
\caption{(Color online) Spatial structures of the SC gap $|\Delta({\tilde {\bf r}})|^2$ (a) and the longitudinal internal magnetic fields normalized by $4\pi C_{\rm GL}$, ${\bf b}_\Delta^{(2)}({\tilde {\bf r}})+{\bf b}_\Delta^{(4)}({\tilde {\bf r}})$, for weak PPB ($\alpha_{\rm M,c}=0.01$) (b) and slightly strong PPB ($\alpha_{\rm M,c}=3.5$) (c) at $t=0.1$ and $H/H_{c2,c}(0)=0.96$ in the case with no AFM fluctuation. $\tilde{{\bf r}}$ denotes the dimensionlesss coordinate defined by $\tilde{{\bf r}}={\bf r}/r_H$. Due to the PPB effect, the magnetic flux is concentrated inside the vortex core as a result of the enhanced magnetic screening outside it, which is reflected in the enhancement of VLFF shown in Fig. \ref{fig:vlff_pauli}. The anisotropy of the internal field around the vortex core is brought by the mismatch between the $d$-wave pairing symmetry and the triangular lattice one \label{fig:magdis}}
\end{center}                                                                                                        
\end{figure}

First, PPB effects on VLFF in the case without AFM fluctuation will be discussed. Figure \ref{fig:vlff_pauli} shows the field dependences of VLFF at $t=0.1$ for $\alpha_{ M,c}=0.01$ (dotted curve), $\alpha_{ M,c}=3.5$ (dashed one), and $\alpha_{ M,c}=5.8$ (solid one). As one can see in Fig. \ref{fig:vlff_pauli}, VLFF is enhanced especially at high fields by the PPB effect. For the sufficiently large $\alpha_{M,c}$, VLFF increases with increasing field while it decreases in the case with the small $\alpha_{M,c}$. The inset in Fig. \ref{fig:vlff_pauli} shows the regular plots of VLFF as a function of magnetic field scaled by $H_{c2,c}(0)$ in the two cases, $\alpha_{ M,c}=0.01$ and $\alpha_{ M,c}=5.8$. The difference in the field dependence of VLFF can be clearly seen. To see that VLFF measures spatial distributions of the internal magnetic field, the longitudinal internal magnetic fields ${\bf b}_\Delta^{(2)}({\bf r})+{\bf b}_\Delta^{(4)}({\bf r})$ in the cases with $\alpha_{\rm M,c}=0.01$ and $\alpha_{\rm M,c}=3.5$ are shown in Fig. \ref{fig:magdis} (b) and (c), respectively. In the SC vortex lattice state shown in Fig. \ref{fig:magdis} (a), the internal field in the direction along the applied magnetic field is induced at the vortex core as a result of the magnetic screening outside the vortex core. The anisotropic structure of ${\bf b}_\Delta^{(2)}({\bf r})+{\bf b}_\Delta^{(4)}({\bf r})$ around the vortex core is due to the mismatch between the four-fold $d$-wave pairing symmetry and the six-fold triangular lattice symmetry. Comparing (b) with (c), one can see that, due to the PPB effect, the magnetic flux is concentrated inside the vortex core as a result of the enhanced magnetic screening outside the vortex core. The enhancement of VLFF shown in Fig. \ref{fig:vlff_pauli} is understood as a consequence of the concentration of the magnetic flux inside the vortex core, or equivalently, the enhanced magnetic screening outside the vortex core \cite{IM}. The increasing behavior in the field dependence of VLFF due to strong PPB has been already argued by Ichioka and Machida who discussed this issue quantitatively by using the quasi-classical approximation where the SC gap and both of the two pair-breaking effects are fully taken into account \cite{IM}. The qualitative consistency of our result with their result indicates that the perturbative expansion with respect to $|\Delta|$ used here is valid in the discussion on VLFF. 
                                                                                                                                                                                                                                                                                                                                                                                                                                                                                                                                                                                                                                                                                                                                                                                                                                                                                                                                                                                                                                                                                                                                                                                                                                                                                                                                                                                                                       

\subsection{AFM fluctuation induced inside the SC phase}
Next, we will show that PPB enhanced sufficiently by increasing the external magnetic field induces AFM fluctuation inside the $d$-wave SC state. 
Since this novel PPB effect has been already studied thoroughly elsewhere \cite{Hatakeyama} in both ${\bf m}\perp {\bf H}$ and ${\bf m}\parallel {\bf H}$, in this subsection, we will only give a brief review of this PPB effect with an example of our result in the case with ${\bf m}\perp {\bf H}$. 

\begin{figure}[t]
\begin{center}
\includegraphics[scale=0.5]{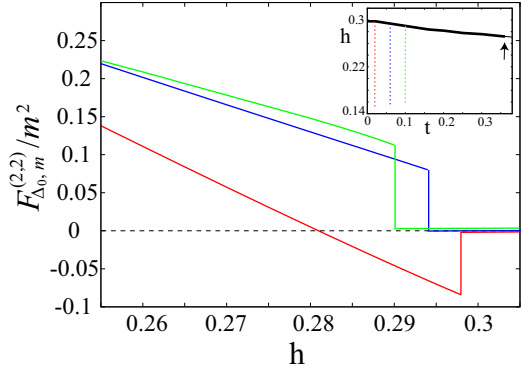}
\caption{(Color online) Field dependences of the SC-AFM coupling term in the free energy ${\cal F}_{\Delta_0,m}^{(2,2)}$ in ${\bf H}\parallel c$ case with ${\bf m}\perp {\bf H}$ at $t=0.02$ [lower (red) curve], $t=0.06$ [middle (blue) one], and $t=0.1$ [upper (green) one]. In obtaining the result, $\alpha_{M,c}=5.8$ and $\delta_{\rm IC}=0.005$ are used. The $H_{c2}(T)$ curve obtained is shown in the inset, where an arrow denotes the point at which the $H_{c2}$ transition changes its nature from the second order to the first one and the FFLO vortex lattice is neglected here. To make it easy to grasp which region in the phase diagram is considered, the temperatures $t=0.02$, $t=0.06$, and $t=0.1$ are indicated by the left (red) vertical line, the center (blue) one, and the right (green) one, respectively in the inset. PPB enhanced strongly by increasing field and decreasing temperature induces AFM fluctuation inside the SC state. The large energy gain in ${\cal F}_{\Delta_0,m}^{(2,2)}$ just below the $H_{c2}$ transition at the low temperature indicates that the AFM fluctuation gets stronger toward $H_{c2}(0)$. \label{fig:AFfluc}}
\end{center}                                                                                                                                                                                                                                                                                                                                                                                                                                                                                                                                                                                                               
\end{figure}

Figure \ref{fig:AFfluc} shows the field dependence of the SC-AFM coupling term ${\cal F}^{(2,2)}_{\Delta_0,m}$ for $\delta_{\rm IC}=0.005$ in ${\bf m}\perp {\bf H}$. Since Eq. (\ref{eq:AF_inst}) determines the AFM instability, negative values of ${\cal F}^{(2,2)}_{\Delta_0,m}$ indicate that an AFM ordering or an enhancement of AFM fluctuation tends to occur in the SC state. Noting the physical implication of ${\cal F}^{(2,2)}_{\Delta_0,m}$, one can see that, due to PPB enhanced by decreasing temperature and increasing field, AFM fluctuation is induced in sufficiently low-temperature and high-field region, while, in higher-temperature and lower-field region, the AFM fluctuation is suppressed by the SC order. The field $h_{CP}$ at which the AFM fluctuation is strongest is fixed at the $H_{c2}$ transition. The jump seen at the $H_{c2}$ transition is due to the finite SC energy gap $|\Delta|$ originating from the first order nature of the $H_{c2}$ transition. These results suggest that the PPB-induced AFM fluctuation gets stronger toward $H_{c2}(0)$ {\it inside} the $d$-wave SC state with finite $|\Delta|$ and that the strong AFM fluctuation with $h_{CP}$ located just below the $H_{c2}$ transition at the extremely low temperature causes the quantum critical phenomena around $H_{c2}(0)$. It should be emphasized that the result obtained here is {\it not} accidental to the GL expansion but intrinsic to the present system. This novel effect of strong PPB has been theoretically confirmed in the Pauli limit case where $|\Delta|$ is fully taken into account \cite{Hatakeyama}. We note that, in ${\bf m}\parallel {\bf H}$, the AFM fluctuation is also induced inside the SC phase by the PPB effect (see Fig. 4 and Fig. 9 in Ref. \cite{Hatakeyama}).
As a typical example of physical phenomena caused by the PPB-induced AFM fluctuation in the SC vortex state, we will discuss effects of the PPB-induced AFM quantum critical fluctuation on VLFF, bearing the experimental data on CeCoIn${}_5$ in our mind.                                                                                                                                                                                                                                                                                                                                                                                                                                                                                                                                                                                                                                                                                                                                                                                                                                                                                                                                                                                                                                                                                                                                                                                                                                                                                                                                                                                                                         
                                                                                                                                                                                                                                                                                                                                                                                                                                                                                                                                                                                                                                                                                                                                                                                                                                                                                                                                                                                                                                                                                                                                                                                                                                                                                                                                                                                                              

\subsection{Effects of AFM fluctuation on VLFF}

\begin{figure}[t]
\begin{center}
\includegraphics[scale=0.35]{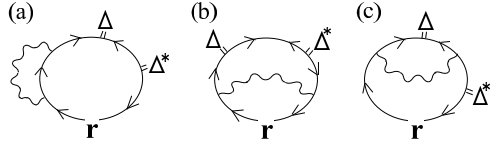}
\caption{Feynman diagrams for the contribution from the AFM fluctuation to the internal magnetic field. The diagrams (a), (b), and (c) are obtained by connecting two wavy lines in the diagrams (d), (e), and (f) in Fig. \ref{fig:Feynman_current}, respectively. A wavy line represents the AFM fluctuation. \label{fig:Feynman_currentAF_fluc}}
\end{center}                                                                                                                                                                                                                                                                                                                      
\end{figure}

Below, we will examine effects of the AFM fluctuation on VLFF. Feynman diagrams for the internal field brought by the AFM fluctuation are shown in Fig. \ref{fig:Feynman_currentAF_fluc} where a wavy line denotes the AFM fluctuation and ${\bf b}^{(0,2)}_m$ has been dropped since it only gives a spatially uniform contribution to the magnetic flux. Then, the spatially varying internal magnetic field is written as ${\bf b}^{(2)}_\Delta({\bf r})+{\bf b}^{(4)}_\Delta({\bf r})+{\bf b}^{(2,2)}_{\rm fluc}({\bf r})$, where
\begin{equation}\label{eq:flux2nd2nd_fluc}                                                                                                                                                                       
\frac{{\bf b}^{(2,2)}_{\rm fluc}({\bf r})}{4\pi C_{\rm GL} } = \frac{8\pi}{7 \zeta (3)} \frac{ r_H }{\xi_0} \frac{|\Delta|^2}{T_c^2}  \,  \sum_{{\bf q}}\frac{\langle | m({\bf q})|^2 \rangle}{T_c^2} \, \sum_{{\bf K}_\perp \neq 0} T^{(2)}_{{\bf K}_\perp} \, \,e^{i{\bf K}_\perp \cdot {\bf r}_\perp} \, \prod _{i=1}^4 \int_0^\infty d\rho_i \sum_{j=1}^{12}  \,p_j  \, \Big\langle |w_{\bf p}|^2 \, \hat{b}^{(2,2)}_{{\bf K}_\perp ,{\bf v}_{\bf p}}(\rho_j;{\bf q},{\bf q}) \, e^{ -\frac{|\eta|^2}{2} (a_j+b_j)^2 } \Big\rangle_{\rm FS}.
\end{equation} 
Here, the commensurate AFM fluctuation ($\delta_{\rm IC}=0$) is assumed for brevity. The properties of the AFM fluctuation are specified by the AFM correlation function $\langle |m({\bf q})|^2 \rangle$ which is introduced phenomenologically through the approximation $\langle |m ({\bf q})|^2 \rangle \simeq T\sum_{\omega_n} \chi ({\bf q},i \omega_n)$ with the familiar form of the susceptibility \cite{Millis, Chubukov_AFM, Pines, Moriya} 
\begin{equation}\label{eq:chi}                                                                                                                      
\chi ({\bf q},i \, \omega_n)= \frac{1}{N(0)}\frac{\xi ^2(h,t) \, \xi_{\rm N}^{-2}}{1+\xi ^2 (h,t)|{\bf q}\times {\hat z}|^2+|\omega_n|/(\Gamma_0 \xi^{-2}(h,t) \, \xi_{\rm N}^{2})},
\end{equation}
where $\xi_{\rm N}$ and $\Gamma_0$ are a length scale and an energy scale characteristic of the AFM fluctuation, respectively, $\xi(h,t)$ is the correlation length which diverges at an AFM instability, and the two dimensional fluctuation is assumed in Eq. (\ref{eq:chi}) based on the experimental data \cite{Curro}. Since, as we discussed in the previous subsection, the AFM fluctuation strongly enhanced at $h_{CP}$ gets stronger with decreasing temperature, we assume that the AFM instability is located at an extremely low temperature and the high field $h_{CP}$. Then, the correlation length is considered to take the form $\xi(h,t)=\xi_{\rm N} \, (t+|1-h/h_{CP}|)^{-1/2}$ \cite{Moriya, Millis_QCP}. We note that the condition $\xi(h,t)^{-2}=0$ corresponds to Eq. (\ref{eq:AF_inst}) which determines the instability of the PPB-induced AFM order. Since $\chi ({\bf q},i \, \omega_n)$ has a dominant contribution at ${\bf q}=0$, we will calculate the right-hand side of Eq. (\ref{eq:flux2nd2nd_fluc}) by using the evaluation $\hat{b}^{(2,2)}_{{\bf K}_\perp ,{\bf v}_{\bf p}}(\rho_j;{\bf q},{\bf q})  \simeq \hat{b}^{(2,2)}_{{\bf K}_\perp ,{\bf v}_{\bf p}}(\rho_j;0,0)$. Then, the part relevant to the summation for ${\bf q}$ in Eq. (\ref{eq:flux2nd2nd_fluc}) is calculated as
\begin{eqnarray}\label{eq:chi_sum}                                                                                                                                                                                                                                                                                                                                                                                                                                                                                    
\sum_{\bf q} \langle |m({\bf q})|^2\rangle 
&\simeq&\frac{q_c}{(2\pi)^2} \int_{-\varepsilon_c}^{\varepsilon_c} d\omega \, \coth \Big(\frac{\omega}{2T}\Big) \int_0^\infty \frac{d(q^2)}{4\pi} \, {\rm Im} \, \chi^R ({\bf q},\omega) \nonumber\\
&=& \frac{1}{N(0) \, \xi_{\rm N}^2 } \frac{q_c \, \varepsilon_c }{(2\pi)^3} \int_0^1 dx \, \coth \Big(\frac{\varepsilon_c}{2T} \, x\Big) \, \tan ^{-1}\Big(\, \frac{\varepsilon_c}{\Gamma_0} \, \frac{\xi^2(h,t)}{\xi_{\rm N}^{2}} \, x\Big),  
\end{eqnarray}
where $\chi^R$ is the retarded susceptibility and the replacement $\sum_{\bf q} \rightarrow \frac{1}{(2\pi)^3}\int dq_z \int d\phi \int qdq$ is used. In the numerical calculation below, ${\varepsilon_c}=2T_c$, $\Gamma_0 = E_F$, $N(0)=p_F^2/(2\pi^2 v_F)$, $q_c=p_F$ are used. We believe that the approximation used here, $\langle |m ({\bf q})|^2 \rangle \simeq T\sum_{\omega_n} \chi ({\bf q},i \omega_n)$, which corresponds to setting $\omega_n \simeq 0$ in the quasiparticle Green's function with an $\omega_n$-dependence, properly gives the AFM contribution since $\chi({\bf q},i \, \omega_n)$ has a dominant contribution at ${\omega_n}=0$. On the other hand, it is known that the $\omega_n$-dependence in the Green's function produces the imaginary part of the self energy of a quasiparticle described by the Feynman diagram (a) in Fig. \ref{fig:Feynman_currentAF_fluc}. Since the imaginary part of the self energy, namely, the quasiparticle damping suppresses the PPB effect \cite{Adachi}, it cannot be neglected in examining any PPB effect. By carrying out the diagrammatic calculation, we can evaluate the imaginary part of the self energy near the Fermi surface as follows:                                      


\begin{equation}\label{eq:self_imag}           
{\rm Im}\, \Sigma^R_\sigma({\bf k}_F,0) \simeq \frac{\pi^2 T}{4} \frac{\xi(t,h)}{p_F \, \xi_{\rm N}^2} \Big(1+\Big[\frac{\xi(h,t)}{2\pi\xi_0} \Big(\frac{\Delta \varepsilon}{T_c}\Big)\Big]^2 \Big)^{-1/2} 
\end{equation}
with $\Delta \varepsilon = -T_c\delta_{\rm IC}+({\overline \sigma}+\sigma)\, \overline{I}T_c$ (for details, see Appendix III). In the expression (\ref{eq:self_imag}), a familiar form of the quasiparticle damping caused by the AFM spin fluctuation near the hot spot \cite{Pines} is extended so that it includes the Zeeman energy shift. Noting that the direction of ${\bf m}$ relative to ${\bf H}$ is reflected in $\Delta \varepsilon$, one can see in Eq. (\ref{eq:self_imag}) that the quasiparticle damping is more effective in ${\bf m}\perp{\bf H}$ than in ${\bf m}\parallel{\bf H}$. In examining effects of the AFM fluctuation, we take the quasiparticle damping into account with the substitution $\varepsilon_n \rightarrow \varepsilon_n+ {\rm sgn}(\varepsilon_n) \, {\rm Im} \,\Sigma_\sigma^R({\bf k}_F,0)$ which is equivalent to the replacement
\begin{equation}\label{eq:AFdamping}                                                                                                                                                                                                                                                                               
\frac{1}{\sinh\big[2\pi t \, x \big]} \rightarrow \frac{\exp\big[-2 \, x \, {\rm Im} \, \Sigma_\sigma^R({\bf k}_F,0)\big]}{\sinh[2\pi t \, x]}
\end{equation}                                                                  
in Eqs. (\ref{eq:f}). In the evaluation of the amplitude of the SC order parameter $\Delta_0$, the replacement (\ref{eq:AFdamping}) is also used.
                                                                                                                                                                                                                                                                                                                                                                                                                                                                                                                                        
\begin{figure}[t]
\begin{center}
\includegraphics[scale=0.5]{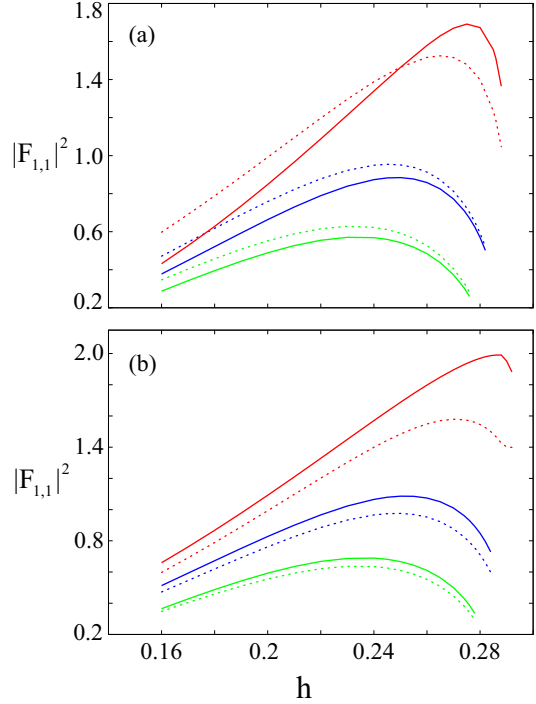}
\caption{(Color online) The $|F_{1,1}(h)|^2$ curves obtained numerically in ${\bf H} \parallel c$ case with ${\bf m}\perp {\bf H}$ (a) and with ${\bf m}\parallel {\bf H}$ (b), where solid (dotted) curves correspond to the case with (without) the additional internal magnetic field brought by the PPB-induced AFM fluctuation. The form factor $|F_{1,1}|^2$ is normalized by $(4\pi C_{\rm GL})^2$. Highest (red), middle (blue), and lowest (green) curves are the $|F_{1,1}(h)|^2$ ones obtained at $t=0.02$, $t=0.06$, and $t=0.1$ in the inset in Fig. \ref{fig:AFfluc}, respectively. Regarding the parameters relevant to the AFM fluctuation, $\xi_{\rm N}=0.6 \xi_0$ with $\xi_0 p_F=7.0$ and $\delta_{\rm IC}=0$ are used and $h_{CP}=0.289$ is assumed so that the location of the AFM instability field is below $H_{c2}(0)$. In the high-field and low-temperature region, VLFF is enhanced by the PPB-induced AFM critical fluctuation. \label{fig:vlffpara-perp}}
\end{center}                                                                                                                                                                                                                                                                                                                                                                                                                                                                                                                                                                                                                                                                                                                                                                                                                       
\end{figure}

\begin{figure}[t]
\begin{center}
\includegraphics[scale=0.65]{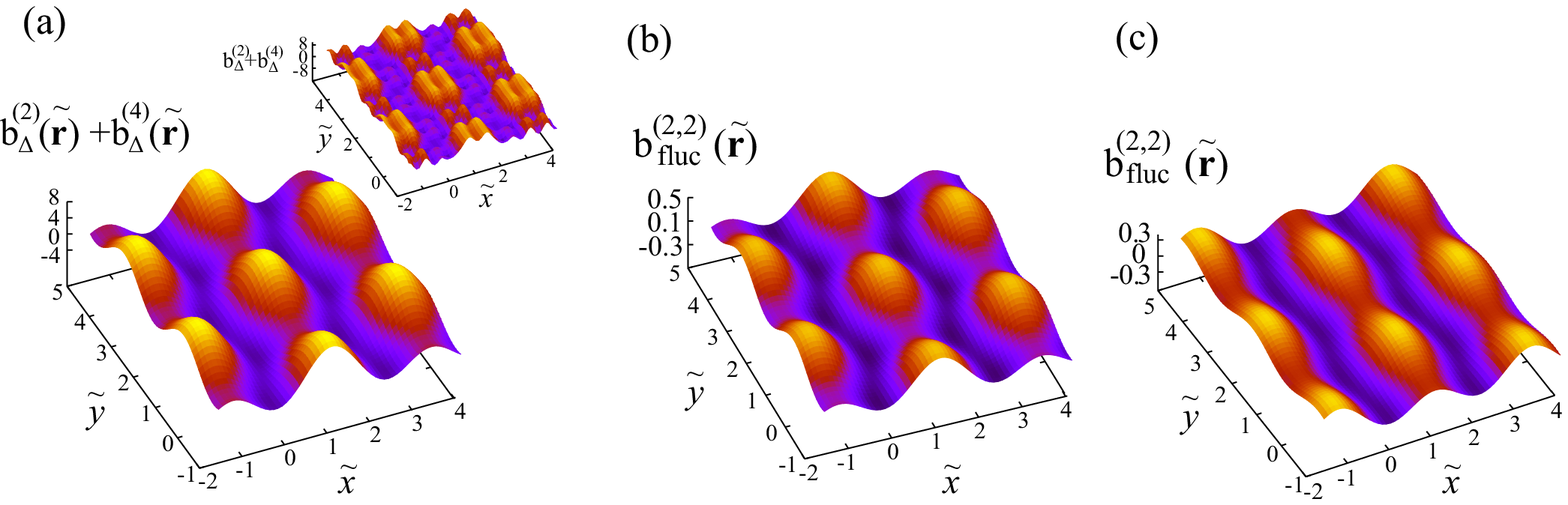}
\caption{(Color online) Spatial distributions of the longitudinal internal magnetic field normalized by $4\pi C_{\rm GL}$ without the AFM contribution ${\bf b}_\Delta^{(2)}({\tilde {\bf r}})+{\bf b}_\Delta^{(4)}({\tilde {\bf r}})$ (a), the internal field brought by the AFM fluctuation ${\bf b}_{\rm fluc}^{(2,2)}({\tilde {\bf r}})$ in ${\bf m}\parallel {\bf H}$ (b), and ${\bf b}_{\rm fluc}^{(2,2)}({\tilde {\bf r}})$ in ${\bf m}\perp {\bf H}$ (c) at $t=0.06$ and $h=0.25$ in the case with $\alpha_{M,c}=5.8$, where dominant contributions coming from the Fourier components $\pm(2,0)$, $\pm(1,1)$, and $\pm(-1,1)$ are extracted from the full internal fields. The inset of (a) shows the full internal field ${\bf b}_\Delta^{(2)}+{\bf b}_\Delta^{(4)}$. Although higher Fourier components become relatively large at low temperatures and express the flux distribution at short length scales, the tendency of the distribution is well described by the six components listed above. Enhancement of the magnetic flux in the vortex core region and its reduction outside the vortex core are commonly seen in both ${\bf b}_\Delta^{(2)}({\tilde {\bf r}})+{\bf b}_\Delta^{(4)}({\tilde {\bf r}})$ and ${\bf b}_{\rm fluc}^{(2,2)}({\tilde {\bf r}})$ in ${\bf m}\parallel {\bf H}$, while, in ${\bf b}_{\rm fluc}^{(2,2)}({\tilde {\bf r}})$ in ${\bf m}\perp {\bf H}$, the magnetic flux is enhanced outside the vortex core. The difference in ${\bf b}_{\rm fluc}^{(2,2)}({\tilde {\bf r}})$ in two cases, ${\bf m}\parallel {\bf H}$ and ${\bf m}\perp {\bf H}$, is reflected in VLFF shown in Fig. \ref{fig:vlffpara-perp}. \label{fig:magAFdis}}
\end{center}                                                                                                                                                                                                                                                                                                                                                                                                                                                                                                                                                                                                                                                                                                                                                                                                                                                                                                                                                                                                                                                                                                                                                                                                                                                                                                                                                                          
\end{figure}

Figure \ref{fig:vlffpara-perp} shows the field dependences of VLFF $|F_{1,1}|^2$ in ${\bf H} \parallel c$ case with ${\bf m}\perp {\bf H}$ (a) and ${\bf m}\parallel {\bf H}$ (b), where solid and dotted curves correspond to the cases with and without the internal magnetic field brought by the PPB-induced AFM fluctuation, respectively, at $t=0.02$ [highest (red) curves], $t=0.06$ [middle (blue) ones], and $t=0.1$ [lowest (green) ones]. The AFM instability field $h_{CP}$ is assumed to be located below $H_{c2}(0)$ so that $h_{CP}$ is consistent with both the experimental result suggestive of the AFM critical point below $H_{c2}(0)$ \cite{Singh, mu_SR, Kasahara} and our theoretical result shown in Fig. \ref{fig:AFfluc}.
In both ${\bf m}\perp {\bf H}$ and ${\bf m}\parallel {\bf H}$, VLFF at the low temperature $t=0.02$ is remarkably enhanced in the high-field range by the PPB-induced AFM critical fluctuation, and as a result, the slope of the $|F_{1,1}(h)|^2$ curve becomes sharper. On the other hand, in ${\bf m}\perp {\bf H}$ (Fig. \ref{fig:vlffpara-perp} (a)), VLFF is rather suppressed by the PPB-induced AFM fluctuation at $t=0.06$, while, in ${\bf m}\parallel {\bf H}$ (Fig. \ref{fig:vlffpara-perp} (b)), the enhancement of VLFF due to the AFM fluctuation can be seen at the same temperature. To understand the physical origin of the enhancement and the suppression of VLFF due to the PPB-induced AFM fluctuation, spatial distributions of the internal field brought by the AFM fluctuation ${\bf b}_{\rm fluc}^{(2,2)}({\bf r})$ in ${\bf m}\parallel {\bf H}$ and ${\bf m}\perp {\bf H}$ at $t=0.06$ and $h=0.25$ are shown in Fig. \ref{fig:magAFdis} (b) and (c), respectively. For comparison, the internal field without the AFM contribution ${\bf b}_\Delta^{(2)}({\bf r})+{\bf b}_\Delta^{(4)}({\bf r})$ at the same temperature and field is also shown in Fig. \ref{fig:magAFdis} (a). In Fig. \ref{fig:magAFdis}, the contributions from the six Fourier components with the shortest $|{\bf K}_\perp|$, $\pm(2,0)$, $\pm(1,1)$, and $\pm(1,-1)$, are extracted from the full summation for ${\bf K}_\perp$ so that one can easily grasp the tendency of the flux distribution. Although, as shown in the inset in (a), ${\bf b}_\Delta^{(2)}({\bf r})+{\bf b}_\Delta^{(4)}({\bf r})$ with the full summation for ${\bf K}_\perp$ has a short length scale structure due to relatively large higher Fourier components, the tendency of the distribution is well described by the six components listed above. As one can see in Fig. \ref{fig:magAFdis}, enhancement of the magnetic flux in the vortex core region and its reduction outside the vortex core are commonly seen in both ${\bf b}_\Delta^{(2)}({\bf r})+{\bf b}_\Delta^{(4)}({\bf r})$ and ${\bf b}_{\rm fluc}^{(2,2)}({\bf r})$ in ${\bf m}\parallel {\bf H}$, while, in ${\bf b}_{\rm fluc}^{(2,2)}({\bf r})$ in ${\bf m}\perp {\bf H}$, the magnetic flux is enhanced outside the vortex core, in other words, the magnetic screening outside the vortex core is suppressed. This result suggests that the enhancement (suppression) of VLFF due to the AFM fluctuation corresponds to the promotion (reduction) of the magnetic screening outside the vortex core. Further, comparing Fig. \ref{fig:vlffpara-perp} (a) with Fig. \ref{fig:AFfluc}, we can see that it depends on a sign of the SC-AFM coupling term ${\cal F}_{\Delta_0,m}^{(2,2)}$ whether the AFM contribution ${\bf b}_{\rm fluc}^{(2,2)}({\bf r})$ promotes the magnetic screening or prevents it. In the high-field and low-temperature region where ${\cal F}_{\Delta_0,m}^{(2,2)}$ is negative, ${\bf b}_{\rm fluc}^{(2,2)}({\bf r})$ enhances VLFF, while, in lower-field and higher-temperature region where ${\cal F}_{\Delta_0,m}^{(2,2)}$ is positive, ${\bf b}_{\rm fluc}^{(2,2)}({\bf r})$ suppresses VLFF. Noting that, from the viewpoint of the SC order, negative values of ${\cal F}^{(2,2)}_{\Delta_0,m}$ indicate that the stability of the SC order is enhanced by the AFM fluctuation, we can understand the mechanism of the enhancement of VLFF due to the PPB-induced AFM fluctuation as follows: the SC state becomes more stable due to the PPB-induced AFM fluctuation, then the SC magnetic screening is promoted, and as a result, VLFF is enhanced by the internal magnetic field originating from the promoted screening current. The falling-down behavior toward the $H_{c2}$ transition in the $|F_{1,1}(h)|^2$ curves is more remarkable in ${\bf m}\perp {\bf H}$ than in ${\bf m}\parallel {\bf H}$, which is due to the fact that the amplitude of the SC energy gap is much suppressed by the stronger quasiparticle damping in ${\bf m}\perp {\bf H}$. 
                                                                                                                                                                                                                                                                                                                                                                                                                                                                                                                                                                                                                                                                                                                                                                                                                                                                                                                                                                                                                                                                                                                                                                                                                                                                                                                                                                                                                                                                                                                                                                                                                                                                                                                                                                                                                                                                                                                                                                                                                                                                                                                                                                                                                                                                                                                                                                                                                                                                                                                                                                                                                                                                                                                                                                                                                                                                                                                                                                                                                                                                                                           
                                                                                                                                                                                                                                                                                                                                                                                                                                                                               
The increasing behavior in the field dependence of VLFF becomes remarkable due to the enhanced magnetic screening caused by the PPB-induced AFM fluctuation. Further, the obtained field dependences of VLFF in lowering temperature in both ${\bf m} \perp {\bf H}$ and ${\bf m} \parallel {\bf H}$, are consistent with the experimental data on CeCoIn${}_5$ \cite{White}. Considering these fact, we could conclude that the strong AFM fluctuation around $H_{c2}(0)$ observed in CeCoIn${}_5$ in ${\bf H} \parallel c$ is the PPB-induced AFM fluctuation enhanced strongly by increasing field and decreasing temperature and that the abrupt increase of VLFF at high fields is a consequence of the magnetic screening enhanced by the PPB-induced AFM critical fluctuation.

Concerning the direction of the AFM moment vector ${\bf m}$, we cannot rule out alternative possibilities of ${\bf m} \perp {\bf H}$ or ${\bf m} \parallel {\bf H}$ by comparing our theoretical result with the experimental data. However, from the experimental data showing that the configuration ${\bf m} \parallel c$ is realized in ${\bf H}\parallel ab$ \cite{Kenzelmann1, Kenzelmann2}, 
it is inferred that ${\bf m}$ is locked along $c$-axis. Further, noting that the AFM fluctuation in the {\it normal} state is suppressed by the Zeeman effect in ${\bf m}\parallel {\bf H}$ while it is not affected by the Zeeman effect in ${\bf m}\perp {\bf H}$, the absence of the AFM order in the SC phase in ${\bf H}\parallel c$ seems to be understood as a result of the suppression of the AFM fluctuation with ${\bf m} \parallel c$ in the normal state. 

\section{Antiferromagnetic order in ${\bf H}\parallel ab$}
In this section, we discuss a possible AFM order realized in vortex states in the $d$-wave superconductor with strong PPB. Throughout this section, we assume that the AFM moment vector is perpendicular to the magnetic field ${\bf m}\perp {\bf H}\parallel ab$, since the configuration ${\bf m} \parallel c$ has been established in the AFM order observed in the HFLT phase of CeCoIn${}_5$ in ${\bf H}\parallel ab$ case.
In the parallel field, strong PPB induces the longitudinal FFLO vortex lattice state in the HFLT corner in the phase diagram \cite{RI07ab} and the FFLO region is large so that it may not be ignored. Although the FFLO state is not taken into account in the analysis on the PPB-induced AFM order in the SC vortex lattice, we will discuss effects of FFLO spatial structures on the AFM order and also on VLFF later.

\subsection{Modulated antiferromagnetic order in the SC vortex lattice}
Below, we consider the situation where the PPB-induced AFM fluctuation becomes so strong that an AFM order appears inside the SC state, i.e., $\big({\cal F}^{(2)}_m+{\cal F}^{(2,2)}_{\Delta_0,m} \big)/m^2 < 0$. Because the SC gap function has the vortex-lattice modulation, an AFM order with a spatial modulation synchronized with the vortex lattice is expected to appear instead of the spatially uniform AFM order. Possible modulated AFM orders are classified into two types: the AFM order localized in the normal-state region, namely, in vortex cores and the one coexistent with the SC order. Since we consider the spatial distribution of the PPB-induced AFM order in the presence of a potential brought by the SC vortex lattice through the SC-AFM coupling term ${\cal F}^{(2,2)}_{\Delta,m}$, it is natural to assume that $m({\bf r})$ takes the following form:
\begin{equation}\label{eq:AFM_trial}
m(X,Y)= \frac{m}{\sqrt{1+ \, w^2 \,\sum' {}_{{\bf K}_\perp \neq 0}|T^{(2)}_{{\bf K}_\perp}|^2}} \bigg(1+ \, w \,\sum _{{\bf K}_\perp  \neq 0} {}' \,T^{(2)}_{{\bf K}_\perp} \exp \Big[i{\bf K}_\perp \cdot (X,Y) \Big] \bigg) ,
\end{equation}
or equivalently,
\begin{equation}
m({\bf q})= \frac{m}{\sqrt{1+ \, w^2 \,\sum' {}_{{\bf K}_\perp \neq 0}|T^{(2)}_{{\bf K}_\perp}|^2}} \bigg(\delta_{{\bf q},0}+  \, w \,T^{(2)}_{{\bf K}_\perp} \, \, \delta_{{\bf q},{\tilde {\bf K}}_\perp \neq 0} \bigg) ,
\end{equation}
where $w$ is a variational parameter to be determined by minimizing $({\cal F}_m^{(2)}+{\cal F}^{(2,2)}_{\Delta_0,m})/m^2$, and $\sum'_{{\bf K}_\perp \neq 0}$ denotes the ${\bf K}_\perp$-summation only for the lowest Fourier components, $(m_1,m_2)=\pm(1,1),\, \pm(1,-1)$, and $\pm(2,0)$. We believe that the restricted summation for ${\bf K}_\perp$ gives correct results because the contribution from higher Fourier components will be negligibly small and will not affect the results. It should be noted that negative (positive) values of $w$ indicate the AFM order enhanced (suppressed) in the vortex core.  

In obtaining spatial structures of AFM orders, we assume that the magnetic flux is uniform. Since, in general, an inhomogeneous internal field is induced by a spatially modulated AFM order, the induced inhomogeneous field may affect the spatial structure of the AFM order. 
In this study, however, we neglect the internal field brought by the AFM order for the following two reasons: First, the leading order AFM contribution ${\bf b}^{(0,2)}_m({\bf r})$ vanishes because of the configuration ${\bf m}\perp {\bf H}$ and does not affect the magnetic flux at all. Second, the internal field brought by the AFM order thorough the SC-AFM coupling ${\bf b}^{(2,2)}_{\Delta,m}({\bf r})$ is considered to be negligibly small since, as we will see later, the transition to the AFM order occurs at a relatively high field so that the amplitude of the AFM order parameter will not grow up to a large value with increasing field.       
                                                        
\begin{figure}[t]
\begin{center}
\includegraphics[scale=0.5]{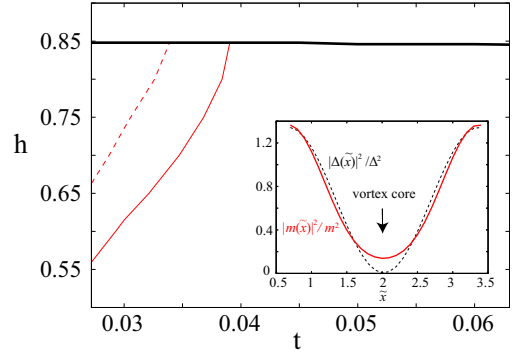}
\caption{(Color online) Example of the numerically obtained temperature-field phase diagram of the PPB-induced AFM order inside the $d$-wave SC vortex lattice state in ${\bf H}\parallel ab$ case, where a thick (black) solid curve and a thin (red) solid one denote the first order $H_{c2}$ transition and the second order transition to the AFM order with a spatial modulation commensurate with the vortex lattice structure, respectively. For comparison, the onset of the spatially uniform AFM order in the SC vortex lattice is also shown as a dashed (red) curve. The parameters used in obtaining the phase diagram are $\alpha_{\rm M,ab}=7.5$, $\delta_{\rm IC}=0.001$, and $T_{\rm N}/T_c=0.027$. The PPB-induced AFM order becomes more stable in the vortex state with the spatial modulation commensurate with the vortex-lattice structure. The inset shows the spatial profiles of the SC gap $|\Delta(\tilde{x})|^2$ (dashed curve) and the modulated AFM order $|m(\tilde{x})|^2$ [solid (red) one] along the trajectory $\tilde{x}$ between neighboring vortex centers. The PPB-induced AFM order realized in the $d$-wave SC vortex state is not localized in the vortex core but coexistent with the SC order. \label{fig:AFSCMab75d0tn0027_0}}
\end{center}                                                                                                                                                                                                                                                                                                                                                                                                                                                                                                                                                                                                                                                            
\end{figure}

\begin{figure}[t]                                                                                                                                                                                                                                                                                                                                                                                                                                                                                                                                                                                                                           
\begin{center}
\includegraphics[scale=0.5]{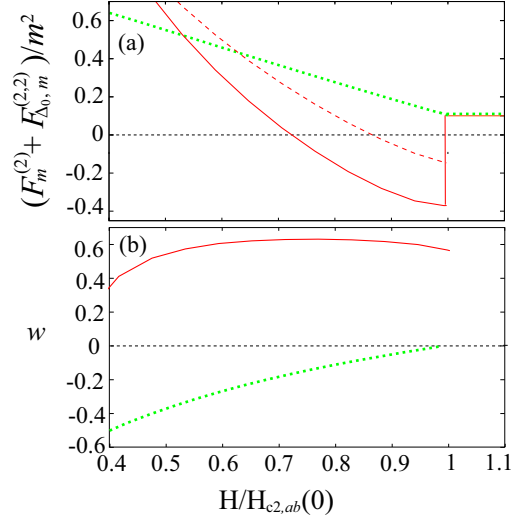}
\caption{(Color online) Field dependences of ${\cal F}_m^{(2)}+{\cal F}_{\Delta_0,m}^{(2,2)}$ (a) and the variational parameter $w$ (b) at $t=0.03$ in Fig. \ref{fig:AFSCMab75d0tn0027_0}. For comparison, results in the case with $\alpha_{M,ab}=0.01$ are also shown in the figures as dotted (green) curves. In (a), one can see that the occurrence of the AFM order in the high-field SC phase lowers the free energy in the case with strong PPB, while, in the case with weak PPB, it never lowers the free energy. Further, as shown in (b), the optimized value of $w$ is positive in the case with strong PPB, suggesting the AFM order coexistent with the SC order, while, in the case with weak PPB, the familiar competition between the two orders is indicated by negative values of $w$. \label{fig:AFmod}}
\end{center}                                                                                                                                                                                                                                                                                                                                                                                                                                                                                                                                                                                                                                                         
\end{figure}

\begin{figure}[t]
\begin{center}
\includegraphics[scale=0.65]{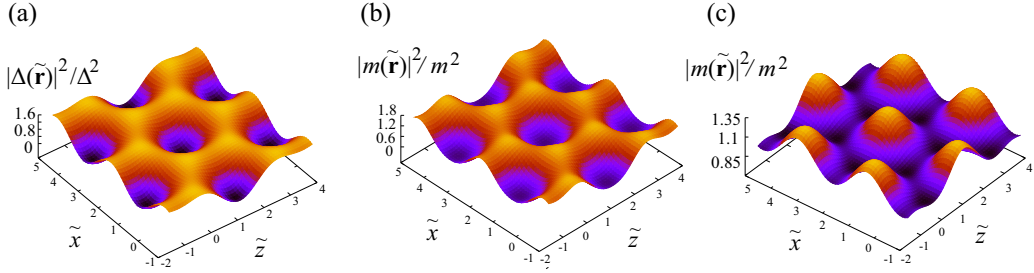}
\caption{(Color online) Spatial structures of the SC gap $|\Delta({\tilde {\bf r}})|^2$ (a) and the modulated AFM order $|m({\tilde {\bf r}})|^2$ in the cases with $\alpha_{M,ab}=7.5$ (b) and $\alpha_{M,ab}=0.01$ (c) at $t=0.03$ and $H/H_{c2,ab}(0)=0.75$ in Fig. \ref{fig:AFmod}. The coordinates $\tilde{x}$ and $\tilde{z}$ are dimensionless and defined as $\tilde{x}=\gamma^{-1/2}x/r_H$ and $\tilde{z}=\gamma^{1/2}z/r_H$, respectively. The figures (b) and (c) correspond to $w=0.63$ and $w=-0.15$, respectively. In the case with strong PPB, the AFM order coexists with the SC order while, in the case with weak PPB, the AFM order competes with the SC order and is localized in the vortex cores.  \label{fig:AFdis}}
\end{center}                                                                                                        
\end{figure}

Figure \ref{fig:AFSCMab75d0tn0027_0} shows an example of the temperature-field phase diagram of the PPB-induced AFM order inside the $d$-wave SC vortex lattice state in ${\bf H}\parallel ab$, where a thin (red) solid curve and a dashed (red) one denote the transition curve to the AFM order with a vortex-lattice modulation and the one to the spatially uniform AFM order, respectively, and both of them are second order transition curves. As one can see in the main panel of Fig. \ref{fig:AFSCMab75d0tn0027_0}, the stability region of the PPB-induced AFM order is expanded by the modulation synchronized with the vortex lattice, and an example of the spatial structure of the modulated AFM order is shown in the inset of Fig. \ref{fig:AFSCMab75d0tn0027_0}. As we will discuss later, the PPB-induced AFM order is {\it not} localized in the vortex core but {\it coexistent} with the SC order. Figures \ref{fig:AFmod} (a) and (b) show the field dependence of ${\cal F}_m^{(2)}+{\cal F}_{\Delta_0,m}^{(2,2)}$ and that of the variational parameter $w$ at $t=0.03$ in Fig. \ref{fig:AFSCMab75d0tn0027_0}, respectively. For comparison, results in the case with $\alpha_{M,ab}=0.01$ are shown in the figures as dotted (green) curves. Figure \ref{fig:AFmod} suggests two important results. First, the AFM order with the vortex-lattice modulation becomes more stable in the vortex state than the uniform one. Second, as clearly seen in Fig. \ref{fig:AFmod} (a), in the case with $\alpha_{M,ab}=7.5$, the AFM order appearing inside the SC state lowers the free energy, while, in the case with $\alpha_{M,ab}=0.01$, it never lowers the free energy. In the case with weak PPB, the conventional competitive nature between SC and AFM orders is reflected in the free energy cost and negative values of $w$, which suggests that, in systems with a sufficiently large $T_{\rm N}/T_{c}$ in zero field, the AFM order localized in the vortex cores, i.e., in the normal-state region where $|\Delta|=0$ can appear \cite{Ogata, Kakuyanagi} as shown in Fig. \ref{fig:AFdis} (c). 
In contrast, as denoted by a solid curve in Fig. \ref{fig:AFmod} (b), optimized values of the variational parameter in the case with strong PPB are positive, indicating the AFM order coexistent with the SC order. An example of the spatial distribution of this PPB-induced AFM order is shown in Fig. \ref{fig:AFdis} (b), where the spatial structure of the vortex lattice is shown in Fig. \ref{fig:AFdis} (a). The AFM order is suppressed in the vortex core and enhanced in the region with large $|\Delta|$. The profiles of the SC gap $|\Delta(x)|^2$ and the PPB-induced AFM order $|m(x)|^2$ along the trajectory $x$ between neighboring vortex centers are shown in the inset of Fig. \ref{fig:AFSCMab75d0tn0027_0}.                                                                                                                                                                                                                                                                                                      

Although we have used a small value for $\delta_{\rm IC}$ in the above numerical calculation, the PPB-induced AFM order becomes much stable in the SC phase as $|\delta_{\rm IC}|$ is increased \cite{IHA}. On the other hand, the incommensurate wave vector observed in the AFM order appearing inside the HFLT phase of CeCoIn${}_5$ \cite{Kenzelmann1, Kenzelmann2} seems to originate from the deviation from the perfect nesting condition of the Fermi surface \cite{Hatakeyama}, which suggests that the deviation $|\delta_{\rm IC}|$ is large enough and the momentum dependence in $|\delta_{\rm IC}|$ should be taken into account in the detailed discussion on the AFM order in the HFLT SC phase.     
                                                                                                                                                                                                                                                                                             
Our result obtained here for the PPB-induced AFM order in the SC {\it vortex lattice} is consistent with recent NMR data suggesting that the AFM order in the HFLT phase of CeCoIn${}_5$ is spatially extended without being localized in the normal-state region \cite{Kumagai_11}. Although the longitudinal FFLO vortex lattice state is ignored in the above calculation, the theoretical study taking account of the longitudinal FFLO modulation along ${\bf H}$ without in-plane vortex-lattice structures included shows that, at least in the high-field side of the FFLO state, the AFM order is not localized in the FFLO nodal planes on which $|\Delta|=0$, but coexistent with the SC order \cite{Hatakeyama}. These results obtained in the two approaches, which are complementary to each other, suggest that the PPB-induced AFM order coexistent with the SC order may be stabilized at least in the high-field side of the longitudinal FFLO vortex lattice state. 

\subsection{VLFF in the modulated AFM order}                                                                                                                                                                                                                                                                                                                                                  
\begin{figure}[t]                                                                                                                                                                                                                                                                                                                                                                                                                                                                                                                                                                                                                                                                                                                                                                                                                                                                                                                                                                                                                                                                                                                                                                                                                
\begin{center}
\includegraphics[scale=0.5]{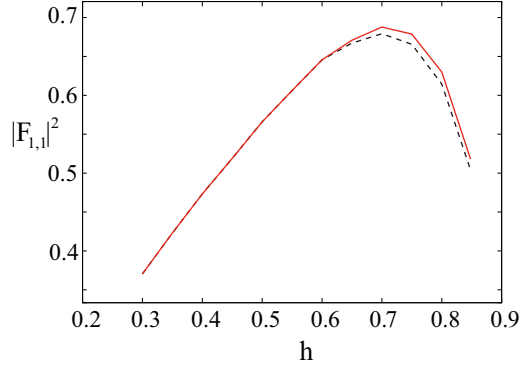}
\caption{(Color online) Field dependence of VLFF $|F_{1,1}|^2$ at $t=0.03$ in Fig. \ref{fig:AFSCMab75d0tn0027_0} obtained in the mean-field approximation, where $|F_{1,1}|^2$ is normalized by $(4\pi C_{\rm GL})^2$. A red solid (dashed) curve corresponds to the result in the presence (absence) of the PPB-induced AFM order with the vortex-lattice modulation. Although VLFF is enhanced by the occurrence of the AFM order, the enhancement is not so remarkable compared with that brought by the PPB-induced AFM fluctuation. The $|F_{1,1}(h)|^2$ curves including the longitudinal FFLO spatial variation of the SC order parameter ignored here are shown in Fig. \ref{fig:magflux_FFLO}. \label{fig:magAForder}}
\end{center}                                                                                                                                      
\end{figure}

We will briefly discuss effects of the PPB-induced AFM {\it order} on VLFF. A red solid (dashed) curve in Fig. \ref{fig:magAForder} denotes the field dependence of VLFF at $t=0.03$ in Fig. \ref{fig:AFSCMab75d0tn0027_0} in the presence (absence) of the PPB-induced AFM order modulating with the vortex lattice. Here, the variational parameter $w$ in Eq.(\ref{eq:AFM_trial}) is determined by minimizing ${\cal F}_{\rm GL}(\Delta_0,m)-{\cal F}_{\rm GL}(\Delta_0,0)$, while the amplitude of the AFM order parameter $|m|$ is obtained by using Eq. (\ref{eq:AFM_amp}). The form factor is enhanced by the PPB-induced AFM order as a result of the additional magnetic screening brought by the occurrence of the PPB-induced AFM order, like in the case only with the PPB-induced AFM {\it fluctuation} examined in Sec. III. However, the enhancement of VLFF brought by the AFM order is quite small compared with that brought by the AFM critical fluctuation. We expect an additional AFM fluctuation around the nonvanishing AFM order parameter, which has not been taken into account in obtaining Fig. \ref{fig:magAForder}, to further enhance VLFF. 

\subsection{Internal magnetic field in the longitudinal FFLO vortex lattice state}

So far, we have not examined effects of the longitudinal FFLO structure of the SC order parameter on the internal field and VLFF which should be seen in the HFLT corner in the phase diagram \cite{RI07ab}. Below, we examine the flux distribution in the longitudinal FFLO vortex lattice although, in turn, an AFM order will be neglected for brevity in contrast to the case of Fig.12. 

\begin{figure}[t]                                                                                                                                                                                                                                                                                                                                                                                                                                                                                                                                                                                                                                                                                                                                                                         
\begin{center}
\includegraphics[scale=0.7]{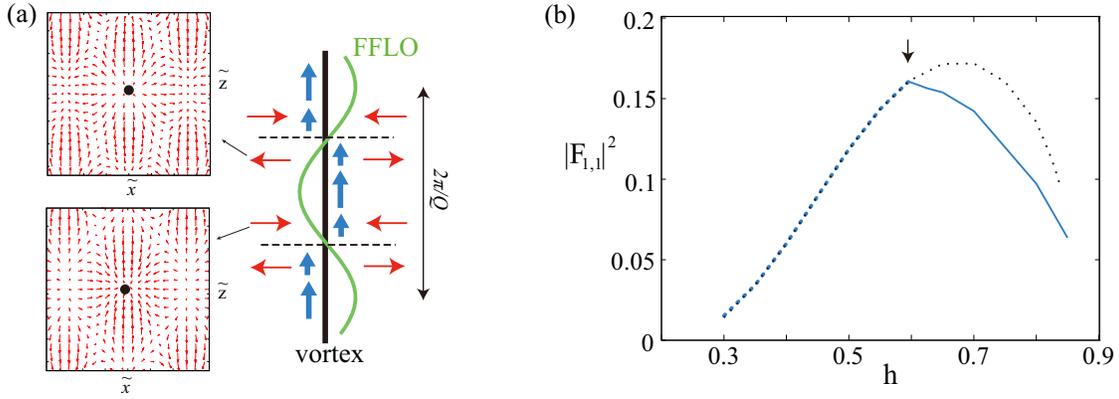}
\caption{(Color online) Spatial flux distribution (a) and the $|F_{1,1}(h)|^2$ curves normalized by $(4\pi C_{\rm GL})^2$ (b) in the longitudinal FFLO vortex lattice state. The AFM order and fluctuation are not taken into account in obtaining these results. The structure of this state is schematically shown in (a) where a solid (black) vertical line, a vertical wavy (green) curve, and horizontal dashed lines indicate a vortex line extending along the applied field, the FFLO modulation, and the nodal planes on which $|\Delta|=0$, respectively. The left upper (lower) panel in (a) shows the numerically obtained transverse field on the surface $QZ=\pi/4$ ($QZ=-\pi/4$) in the FFLO state with $2\pi \xi_0 Q =0.55$ appearing at $t=0.1$ and $h=0.65$, where a (black) dot denotes the vortex center. An outgoing field from the vortex center is seen at $QZ=\pi/4$, while, on the surface $QZ=-\pi/4$, rather an incoming field is induced. Correspondingly, the longitudinal magnetic flux at the vortex core denoted by vertical (blue) arrows varies along the applied field with its minimum at the nodal plane and maximum at the surface on which $|\Delta|$ is the largest.  In (b), (blue) solid and (blue and black) dotted curves denote the $|F_{1,1}(h)|^2$ ones with and without the longitudinal FFLO modulation at $t=0.1$, respectively, and the arrow denotes the corresponding second order transition field to the longitudinal FFLO vortex state. The VLFF is suppressed due to the FFLO spatial modulation of the SC order parameter. \label{fig:magflux_FFLO}}
\end{center}                                                                                                                                                                                                                                                                                                                                                                                                                                                                                                                                                                                     
\end{figure}

In the longitudinal FFLO vortex lattice state, the SC gap function is expressed in the form 
\begin{equation}\label{eq:FFLO}
\Delta({\bf r})=\Delta \varphi_0(X,Y) \, \sqrt{2} \cos\big(Q \,Z \big)
\end{equation}
with the modulation wave vector $Q {\hat Z}$ parallel to the magnetic field. This FFLO modulation in the SC order parameter affects the internal field through the operator acting on the SC gap function $\exp\big[i A {\bf v}_{\bf p}\cdot {\bf \Pi}\big]\Delta({\bf r})=\sqrt{2} \Delta \exp\big[i A {\bf v}_{\bf p}\cdot {\bf \Pi}_\perp \big]\varphi_0(X,Y) \, \exp\big[i A \big({\bf v}_{\bf p}\cdot \hat{Z} \big) (-i\,\nabla_Z )\big]\cos\big(Q \,Z \big)$, and then, the Maxwell equation $-\nabla \times {\bf b}_\Delta^{(2)}({\bf r})/(4\pi C_{\rm GL})=\nabla \times \big(\sum_{{\bf K}_\perp \neq 0}\big[m_{{\bf K}_\perp,Q}^{(+)} + m_{{\bf K}_\perp,Q}^{(-)} \cos\big(2\, Q\, Z\big)\big]e^{i\, {\bf K}_\perp \cdot (X,Y)} \big) \hat{Z}$ is obtained. It can be shown that the Maxwell equation has a solution with a transverse component in ${\bf b}_\Delta^{(2)}({\bf r})$ \cite{Houzet}. The resulting O($|\Delta|^2$) contribution to the internal field is expressed in the form 
\begin{equation}\label{eq:flux2nd_FFLO}          
\frac{-{\bf b}_\Delta^{(2)}({\bf r})}{4\pi C_{\rm GL}} 
=\hat{Z} \sum_{{\bf K}_\perp \neq 0}\Big[m_{{\bf K}_\perp,Q}^{(+)} + \frac{|{\tilde {\bf K}}_\perp|^2 \, m_{{\bf K}_\perp,Q}^{(-)}}{|{\tilde {\bf K}}_\perp|^2+(2Q)^2} \cos\big(2\, Q\, Z\big)\Big]e^{i\, {\bf K}_\perp \cdot (X,Y)} -\sum_{{\bf K}_\perp \neq 0} \frac{2\,i\,Q \, {\tilde {\bf K}}_\perp\, m_{{\bf K}_\perp,Q}^{(-)}}{|{\tilde {\bf K}}_\perp|^2+(2Q)^2} \sin\big(2\, Q\, Z\big) e^{i\, {\bf K}_\perp \cdot (X,Y)},
\end{equation}
where
\begin{equation}
m_{{\bf K}_\perp,Q}^{(\pm)}= \frac{8\pi}{7 \zeta (3)} \frac{r_H}{\xi_0}  \frac{|\Delta|^2}{T_c^2} \int_0^\infty d\rho_1 \, d\rho_2 \, T^{(2)}_{{\bf K}_\perp}  \Big\langle |w_{\bf p}|^2 \, \hat{b}^{(2)}_{{\bf K}_\perp,{\bf v}_{\bf p}}(\rho_{1,2}) \, e^{-\frac{|\eta|^2}{2} (\rho_1+\rho_2)^2} \, \cos\big(Q \, \big( {\bf v}_{\bf p} \cdot \hat{Z} \big)  (\rho_1\pm \rho_2) \big)\Big\rangle_{\rm FS} .
\end{equation}                                                                                                                                                                                                     
As one can see in Eq. (\ref{eq:flux2nd_FFLO}), the transverse magnetic field is induced by the finite FFLO modulation ($Q\neq 0$) and the longitudinal magnetic flux varies along the FFLO modulation. An example of the flux distribution in the longitudinal FFLO vortex lattice is shown in Fig. \ref{fig:magflux_FFLO} (a). One can see that outgoing and incoming fields are induced and correspondingly, the longitudinal magnetic flux varies along the FFLO modulation. 
The experimentally measured VLFF is expected to reflect the internal field which is spatially averaged over the range $[-\pi/Q,\pi/Q]$ in the $Z$ direction. Then, the transverse incoming and outgoing fields cancel each other. The $|\Delta|^4$ contribution to the spatially averaged internal magnetic field is given by replacing $\hat{b}^{(4)}_{{\bf K}_\perp,{\bf v}_{\bf p}}(\rho_j) $ in Eq. (\ref{eq:flux4th}) by $\hat{b}^{(4)}_{{\bf K}_\perp,{\bf v}_{\bf p}} (\rho_j)\,  \big[ \cos \big(Q \, \big( {\bf v}_{\bf p} \cdot \hat{Z} \big) \, (\alpha_j-\beta_j-\gamma_j+\delta_j) \big) \, /2 + \cos \big(Q \, \big( {\bf v}_{\bf p} \cdot \hat{Z} \big) (\alpha_j+\beta_j)\big) \, \cos\big(Q \, \big( {\bf v}_{\bf p} \cdot \hat{Z} \big) (\gamma_j+\delta_j)\big) \big]$. Figure \ref{fig:magflux_FFLO} (b) shows typical $|F_{1,1}(h)|^2$ curves in the longitudinal FFLO vortex lattice without an AFM order, where solid (blue) and dotted curves correspond to the results with and without the FFLO longitudinal modulation, respectively, and the arrow denotes the FFLO transition field. 
In obtaining the result, the modulation $Q$ is determined selfconsistently from the GL free energy functional \cite{RI07ab}, and further, it is confirmed that the FFLO transition is of second order. As one can see in the figure, VLFF is suppressed by the longitudinal FFLO modulation and the slope of the $|F_{1,1}(h)|^2$ curve changes its sign quite close to the FFLO transition field. An abrupt falling down behavior in VLFF has been experimentally observed near the transition between the low field SC phase and the HFLT one in ${\bf H}\parallel ab$ case \cite{White_Phd}, which seems to imply the formation of the longitudinal FFLO vortex lattice state in the HFLT phase. On the other hand, an AFM order and corresponding AFM fluctuation which are not taken into account in the above argument about the FFLO state may affect the result. It should be noted, however, that the quasiparticle damping brought by the AFM critical fluctuation should suppress the PPB effect so that the onset of the FFLO transition field in the above calculation should be shifted to a higher field and that the FFLO region should be narrower. In order to investigate the high-field and low-temperature SC phase in ${\bf H}\parallel ab$, further studies taking account of the longitudinal FFLO vortex lattice, the PPB-induced AFM order, and fluctuation around the AFM transition consistently will be needed. 

In the last of this subsection, we will discuss effects of the longitudinal FFLO modulation on VLFF {\it in ${\bf H}\parallel c$ case}. The HFLT phase appearing just below the $H_{c2}(T)$ curve as a narrow region in ${\bf H}\parallel c$ case \cite{Bianchi03, Kumagai_06} has been considered to be a realization of the longitudinal FFLO vortex lattice \cite{RI07c}. On the other hand, VLFF obtained in the neutron scattering experiment does not show any structure around the transition between the HFLT phase and the lower field phase \cite{White}, in contrast to the abrupt falling down behavior in ${\bf H}\parallel ab$ case discussed above. The absence of a precursor of the FFLO state in VLFF in ${\bf H}\parallel c$ case may be due to the fact that the effect of the FFLO modulation on VLFF is too small just above the FFLO transition to be seen. Although VLFF should be suppressed by the FFLO modulation $Q$, drastic suppression cannot be seen because, in ${\bf H}\parallel c$ case, the stability region of the FFLO state is quite narrow and $Q$ cannot grow up to a large value. We believe that, if VLFF is detected in sufficiently high resolution, the onset of the FFLO transition should be observed in VLFF in ${\bf H}\parallel c$ case. 
                                                                                                                                                                                                                                                                                                                                                                                                                                                                                                                                                                                                                                                                                                                                                                                                                                                                  

\section{Summary}
Antiferromagnetic (AFM) fluctuation and order induced inside a $d$-wave superconducting (SC) phase by strong Pauli-paramagnetic pair-breaking (PPB) \cite{IHA} have been theoretically studied, focusing on how they are affected by spatial modulations of a SC vortex lattice. Based on the microscopic calculation taking account of both orbital and paramagnetic pair-breaking effects, we have shown that, with increasing field and decreasing temperature, the SC vortex lattice state becomes more stable by inducing AFM fluctuation inside it: From the viewpoint of the SC order, the magnetic screening is promoted by the induced AFM fluctuation in the high-field SC phase. In other words, the AFM order can more easily appear inside the SC vortex lattice state than in the normal state. These implications have been concretely discussed in connection with the anomalous SC phenomena observed in CeCoIn${}_5$. First, in ${\bf H}\parallel c$ case where the AFM quantum critical behavior around $H_{c2}(0)$ is experimentally observed, such appearance of the AFM fluctuation in the SC state in higher fields has been explained as a result of the PPB enhanced with increasing the field. Then, it is found that the vortex lattice form factor (VLFF) is enhanced by the additional magnetic screening brought by the coupling between the induced AFM fluctuation and the spatial modulation of the SC vortex lattice and that the VLFF's enhancement becomes much remarkable as the AFM instability is approached. The obtained temperature and field dependences of VLFF are consistent with the experimental result \cite{VLFF, White}, which suggests the anomalous field dependence of VLFF observed in the experiments is a consequence of the PPB-induced AFM critical fluctuation. Second, in ${\bf H}\parallel ab$ case where an AFM order with ${\bf m}\perp {\bf H}$ appears inside the SC phase in CeCoIn${}_5$, it is found that the vortex-lattice modulation promotes the AFM fluctuation and that, as a result, a spatially modulated AFM order synchronized with the vortex lattice is stabilized in the SC vortex state. In contrast to the conventional competitive nature between the AFM and SC orders, the PPB-induced AFM order in the vortex lattice coexists with the nonvanishing SC order parameter and does not localize in the vortex cores. 
The spatial modulation peculiar to the longitudinal FFLO state, which seems to be realized in the HFLT phase of CeCoIn${}_5$ \cite{Kumagai_11}, has been neglected in our analysis on the AFM order. According to the theoretical study taking account of the longitudinal FFLO modulation in the Pauli limit without vortex lattice structures included \cite{Hatakeyama}, a possible PPB-induced AFM order in the high-field side of the FFLO state is not localized in the FFLO nodal plane but prefers to coexist with the nonvanishing SC order parameter. The unconventional coexistence of the AFM and SC orders, obtained in the two approaches complementary to each other, is consistent with a picture on the HFLT phase of CeCoIn$_5$ suggested from the recent NMR data \cite{Kumagai_11}. Further, our study on effects of the longitudinal FFLO spatial modulation and the AFM order on VLFF performed additionally has shown that, as well as the AFM fluctuation, the AFM order enhances VLFF, while the longitudinal FFLO modulation leads to a significant reduction of VLFF similar to that seen in CeCoIn${}_5$. These results on the magnetic properties in high field superconductors with strong PPB imply that the HFLT state of CeCoIn${}_5$ is a coupled phase composed of a PPB-induced AFM order and the $d$-wave SC order accompanied by the longitudinal FFLO modulation. 
                                                                                                                                                                                                                                                                                                                                                                                                                                                                                                                                                                                                                                                                                                                                                                                                                                                                                                                                                                                                                                                                                                                                                                                                                                                                                                                                                                                                                                                                                                                                                                                                                                                                                                                                                                                                                                                                                                                                                                                                                                                                                                                                                                                                                                                                                                                                                                                                                                                                                                                                                                                                                                                                                                                                                         
\section{acknowledgement}
We are grateful to Y. Matsuda, K. Kumagai, T. Shibauchi, M. Eskildsen, M. Kenzelmann, S. Gerber, and M. Sigrist for usefufl discussions and A. Bianchi for providing us with a copy of Ref. \cite{White_Phd} and discussion on the vortex lattice in the HFLT phase. This work was partly supported by Grant-in-Aid for Scientific Research [No. 21540360] from JSPS, Japan.

\section{Appendix}
\subsection{Orbital pair-breaking effect}
Here, we show the details of the calculation in obtaining Eq. (\ref{eq:orbital_operation}) and derive equations useful in incorporating the orbital pair-breaking effect. In the case with ${\bf H} \parallel {\hat z}$ (${\bf H} \parallel {\hat y}$), the vector potential can be expressed as ${\bf A}_0=(0,Hx,0)$ (${\bf A}_0=(Hz,0,0)$), and then, $[\Pi_x,\Pi_y]=-ir_H^{-2}$ ($[\Pi_z,\Pi_x]=-ir_H^{-2}$) is satisfied.
In ${\bf H} \parallel {\hat z}$ case, by introducing the operators $\Pi_\pm = r_H(\Pi_x \pm i \Pi_y)/{\sqrt 2}$, we obtain
\begin{eqnarray}\label{eq:orbital_op}
\exp\Big[iA\, T_c^{-1}{\bf v}_{\bf p}\cdot {\bf \Pi} \Big]&=&e^{-\frac{|\eta|^2}{2}A^2 } \, e^{ iA\,  \eta^\ast \Pi_+} \, e^{i A \, \eta \Pi_-}, \nonumber\\
e^{ iA\,  \eta^\ast \Pi_+}&=& e^{ (iA \eta^\ast r_H/2)^2 [\Pi_x,i\Pi_y]} \, e^{- (A \, \eta^\ast  r_H /{\sqrt 2}) \, \Pi_y} \, e^{ i (A \, \eta^\ast r_H/{\sqrt 2}) \, \Pi_x},
\end{eqnarray}
where the operator identity $e^{[\hat{A},\hat{B}]}=e^{-[\hat{A},\hat{B}]/2} \, e^{\hat{A}} \, e^{\hat{B}}$ is used. In ${\bf H} \parallel {\hat y}$ case, $(\Pi_x,\Pi_y)$ is replaced by $(\gamma^{-1/2} \Pi_z,\gamma^{1/2} \Pi_x)$ in Eq. (\ref{eq:orbital_op}). Equation (\ref{eq:orbital_operation}) is obtained by using Eq. (\ref{eq:orbital_op}).
Since the Fourier transformation of $\varphi_0(X+\alpha r_H,Y)\varphi_0^\ast(X+\beta r_H,Y)$ with the reciprocal lattice vector ${\bf K}_\perp=(K_x,K_y)$ is calculated as 
\begin{eqnarray}\label{eq:orbital_operation_FT}
&&\int \frac{dX \, dY}{L_X L_Y} \, \varphi_0(X+\alpha r_H,Y)\varphi_0^\ast(X+\beta r_H,Y)e^{-i{\bf K}_\perp \cdot (X,Y)} =T_{{\bf K}_\perp}^{(2)} \, e^{-\frac{1}{4}(\alpha-\beta)^2} \exp\Big[i\frac{K_x r_H}{2}(\alpha+\beta)\Big] \, \exp\Big[-\frac{K_y r_H}{2}(\alpha-\beta)\Big], \nonumber\\
&&T_{{\bf K}_\perp}^{(2)}=\delta_{\frac{K_y r_H}{k}+\frac{k}{\pi}K_x r_H,even}|_{\frac{K_y r_H}{k}:integer} \, \cos\bigg(\frac{\pi}{2}\frac{K_y r_H}{k}\Big[\frac{K_y r_H}{k}+\frac{k}{\pi}K_x r_H\Big]\bigg) \,\exp\Big[-\frac{1}{4}|{\bf K}_\perp r_H|^2\Big] ,
\end{eqnarray} 
we obtain 
\begin{eqnarray}\label{eq:formula2nd}
&&\exp \bigg[is_{\varepsilon_n} T_c^{-1} \Big({\bf v}_{\bf p}\cdot \Pi_1 \, \alpha + {\bf v}_{\bf p}\cdot \Pi^\dagger_2 \, \beta \Big)\bigg] \, \varphi_0 ({\bf s}_1) \varphi_0^\ast ({\bf s}_2) |_{{\bf s}_i \rightarrow (X,Y)} \nonumber\\
&&=e^{-\frac{1}{2} \big[ \alpha ^2 (|\eta|^2-\eta^{\ast \, 2})+\beta^2 (|\eta| -\eta^2)\big]} \varphi_0(X+s_\varepsilon \alpha \eta^\ast {\sqrt 2} r_H,Y)  \varphi_0^\ast(X-s_\varepsilon \beta \eta {\sqrt 2} r_H,Y)\nonumber\\
&&=\exp\Big[-\frac{|\eta|^2}{2}  (\alpha+\beta)^2\Big] \,\sum_{{\bf K}_\perp} T^{(2)}_{{\bf K}_\perp} \,e^{i{\bf K}_\perp \cdot (X,Y)} \, \exp \bigg[ i s_{\varepsilon_n}  \Big(\frac{({\mbox{\boldmath $\eta$}}\cdot {\bf K}_\perp r_H )}{{\sqrt 2}}(\alpha-\beta) + i \frac{({\mbox{\boldmath $\eta$}}\times {\bf K}_\perp r_H)_z}{{\sqrt 2}}(\alpha+\beta)\Big) \bigg]. 
\end{eqnarray}
In the same manner, the identity corresponding to the fourth order term is given by
\begin{eqnarray}\label{eq:formula4th}
&&{\rm exp}\bigg[ i \, s_{\varepsilon_n} T_c^{-1}\Big({\bf v}_{\bf p}\cdot{\bf \Pi}_1 \, \alpha + {\bf v}_{\bf p}\cdot{\bf \Pi}^\dagger_2 \, \beta + {\bf v}_{\bf p}\cdot{\bf \Pi}_3 \, \gamma + {\bf v}_{\bf p}\cdot{\bf \Pi}^\dagger_4 \, \delta \Big)\bigg] \, \varphi_0({\bf s}_1) \varphi_0^\ast({\bf s}_2)\varphi({\bf s}_3) \varphi_0^\ast({\bf s}_4) |_{{\bf s}_i \rightarrow (X,Y)} \nonumber\\
&=& {\rm exp}\bigg[ -\frac{|\eta|^2}{2} \Big( \alpha ^2 +\beta ^2 +\gamma ^2 +\delta ^2 +(\alpha+\gamma)(\beta+\delta) \Big) \bigg] \, {\rm exp}\bigg[ -\frac{1}{4}\Big( (\alpha -\gamma )^2 \eta^{\ast \,  2} + (\beta -\delta )^2 \eta ^2 \Big) \bigg] \nonumber\\
&&\times \frac{k} {\sqrt{2\pi}}\sum_{{\bf K}_\perp} T^{(4)}_{{\bf K}_\perp} \, e^{i {{\bf K}_\perp}\cdot (X,Y)} \sum_{l_1,l_2= -\infty}^\infty T^{(4)}(l_1,l_2,{\bf K}_\perp) \nonumber\\
&&\times {\rm exp}\bigg[ i s_{\varepsilon_n} \Big(-i\eta^\ast \frac{k}{\sqrt 2} (l_1-l_2)(\alpha-\gamma) - i \eta\frac{k}{\sqrt 2}(l_1+l_2+\frac{K_y r_H}{k})(\beta-\delta) \Big) \bigg] \nonumber\\
&&\times {\rm exp}\bigg[ i s_{\varepsilon_n} \Big(\frac{({\mbox{\boldmath $\eta$}}\cdot {\bf K}_\perp r_H)}{2{\sqrt 2}} \big\{(\alpha+\gamma)-(\beta+\delta) \big\} + i \frac{({\mbox{\boldmath $\eta$}}\times {\bf K}_\perp r_H)_z}{2{\sqrt 2}} \big\{ (\alpha + \gamma) + (\beta+\delta) \big\} \Big) \bigg], \nonumber\\
&&T^{(4)}_{{\bf K}_\perp}=\delta_{\frac{k}{\pi}K_x r_H + \frac{K_y r_H}{k}, {\rm even}}|_{\frac{K_y r_H}{k}:{\rm integer}} \, \exp\bigg[-i \, \frac{\pi}{4}\frac{K_y r_H}{k}\Big(\frac{k}{\pi}K_x r_H + \frac{K_y r_H}{k} \Big)\bigg] \,\exp\Big[-\frac{1}{8}|{\bf K}_\perp r_H|^2\Big], \nonumber\\
&&T^{(4)}(l_1,l_2,{\bf K}_\perp) = \cos\Big( \frac{\pi}{2}(\frac{k}{\pi}K_x r_H + \frac{K_y r_H}{k})(l_1+l_2)\Big) \nonumber\\
&& \qquad \qquad \qquad \quad \times {\rm exp}\bigg[ i\frac{\pi}{4}\Big( (l_1-l_2)^2 - (l_1+l_2+\frac{K_y r_H}{k})^2 \Big) \bigg] \, {\rm exp}\bigg[ -\frac{k^2}{4}\Big( (l_1-l_2)^2 + (l_1+l_2+\frac{K_y r_H}{k})^2 \Big) \bigg],
\end{eqnarray}
where $l_1$ and $l_2$ are integers.

\subsection{GL coefficients $V_i$ in Eq. (\ref{eq:F_s})}
Since the GL coefficients $V_i$ in Eq. (\ref{eq:F_s}) have been already derived elsewhere \cite{Adachi, RI07ab}, we only show the final expressions of $V_i$ which are given by
\begin{eqnarray}\label{eq:GLcoefficient}
V_2&=&\frac{1}{2}\ln(h)+\int_0^\infty d\rho\, \bigg[\frac{1}{\rho}\exp\Big(-\frac{\pi^2\xi_0^2}{r_H^2}\, \rho^2\Big)-f_{\cos} (\rho,\rho) \Big\langle |w_{\bf p}|^2 \exp\Big( -\frac{|\eta|^2}{2}\, \rho^2\Big)\Big\rangle_{\rm FS} \bigg], \nonumber\\
V_4=&=&2\prod_{i=1}^3\int_0^\infty d\rho_i \, f_{\cos} \big(\sum_{i=1}^3 \rho_i,\sum_{i=1}^3 \rho_i \big) \Big\langle |w_{\bf p}|^4 \exp\bigg[-\frac{1}{2}\Big( -\frac{1}{2}R_{24} + R_{14}\Big)\bigg]\cos(I_4) \Big\rangle_{\rm FS}, \nonumber\\
V_6&=&-6\prod_{i=1}^5\int_0^\infty d\rho_i \, f_{\cos} \big(\sum_{i=1}^5 \rho_i,\sum_{i=1}^5 \rho_i \big) \Big\langle |w_{\bf p}|^6 \exp\bigg[-\frac{1}{2}\Big( R_{16} + R_{26}\Big)\bigg]\cos(I_6) \Big\rangle_{\rm FS},
\end{eqnarray}                                                                                                                                            
where
\begin{eqnarray}
R_{14}&=& |\eta|^2\Big(\sum_{i=1}^3 \rho_i^2 + \rho_2(\rho_3+\rho_1) \Big), \nonumber\\
R_{24}&=& {\rm Re}(\eta^2)[\rho_2^2+(\rho_3-\rho_1)^2], \nonumber\\
I_4&=& \frac{{\rm Im}(\eta^2)}{4}[\rho_2^2-(\rho_3-\rho_1)^2], \nonumber\\
R_{16}&=& |\eta|^2 \Big(e_1+e_2+e_3+\frac{2}{3}e_4 e_5\Big), \nonumber\\
R_{26}&=& {\rm Re}(\eta^2)\Big(e_1+e_2+e_3-\frac{e_4^2+e_5^2}{3}-\frac{2}{3}(e_6 + e_7 + e_8 + e_9)\Big), \nonumber\\
I_6&=& \frac{{\rm Im}(\eta^2)}{4}\Big(e_1+e_2-e_3-\frac{e_4^2-e_5^2}{3}-\frac{2}{3}(e_6 + e_7 - e_8 - e_9)\Big), \nonumber\\
e_1&=& (\rho_3+\rho_5)^2+(\rho_3+\rho_4)^2, \nonumber\\
e_2&=& (\rho_1+\rho_4+\rho_5)^2, \nonumber\\
e_3&=& \rho_3^2+\rho_4^2+ (\rho_2-\rho_5)^2, \nonumber\\
e_4&=& \rho_1+ 2(\rho_3+\rho_4+\rho_5), \nonumber\\
e_5&=& \rho_2-\rho_3-\rho_4-\rho_5 , \nonumber\\
e_6&=& (\rho_4-\rho_5)^2+(\rho_1+\rho_5-\rho_3)^2, \nonumber\\
e_7&=& (\rho_1+\rho_4-\rho_3)^2, \nonumber\\
e_8&=& (\rho_3-\rho_4)^2+(\rho_2+\rho_3-\rho_5)^2, \nonumber\\
e_9&=& (\rho_2+\rho_4-\rho_5)^2 .
\end{eqnarray}

\subsection{Imaginary part of the self energy }
The self energy of a quasiparticle is expressed as \cite{Chubukov_AFM}
\begin{eqnarray}           
\Sigma_\sigma({\bf k},i\varepsilon_n) &=& T\sum_{\varepsilon_1}\sum_{\bf q}{\cal G}_{\varepsilon_1,{\overline \sigma}}({\bf k}+{\bf Q}_0-{\bf q}) \, \chi({\bf q},i(\varepsilon_n-\varepsilon_1)).
\end{eqnarray}
After carrying out an analytic continuation, we obtain the imaginary part of the self energy on the Fermi surface as follows:
\begin{eqnarray}\label{eq:AFimag_appendix1}
{\rm Im} \, \Sigma^R_\sigma({\bf k},0)&=&\frac{1}{ 2 \pi}\sum_{\bf q}\int_{-\infty}^{\infty}dx \, {\rm Im}\, {\cal G}^R_{-i \, x,{\overline \sigma}}({\bf k}+{\bf Q}_0-{\bf q}) \,{\rm Im}\, \chi^R({\bf q},x)  \Big[\coth\Big( \frac{x}{2T}\Big)-\tanh\Big( \frac{x}{2T}\Big) \Big] \nonumber\\
&\simeq&\frac{q_c}{ 4(2 \pi)^2 } \int \frac{d\phi}{2\pi} \int d(q^2) \, \delta \Big(\varepsilon({\bf k}+{\bf Q}_0+{\bf q})+{\overline \sigma}\overline{I}\,T_c \Big) \, \frac{\xi ^2(h,t)}{N(0)\xi_{\rm N}^2} \frac{2T}{1+\xi ^2 (h,t) \, q^2},
\end{eqnarray}                                                                                                                                                                                             
where $\Sigma^R$ is the retarded self energy, ${\cal G}^R$ is the retarded Green's function, and the inequality $|x|/2T \ll 1$ is used. Since the $\delta$-function in Eq. (\ref{eq:AFimag_appendix1}) is reduced to
\begin{eqnarray}\label{eq:AFimag_appendix2}
\delta \Big(T_c\delta_{\rm IC}+{\bf v}_{\bf k}\cdot{\bf q}-({\overline \sigma}+\sigma)\overline{I} \,T_c \Big)&
=&\frac{\theta( q^2 -[\Delta \varepsilon /v_F]^2)}{q \, v_F \, |\sin\phi_0|}\delta(\phi-\phi_0)
\end{eqnarray}   
with $\phi_0=\cos^{-1}\big(  \Delta\varepsilon /(q \, v_F ) \big)$, we obtain             
\begin{eqnarray}\label{eq:self_imag0}           
{\rm Im}\, \Sigma^R_\sigma({\bf k}_F,0)&=&\frac{\pi T}{ 2 }\frac{1}{N(0)\xi_{\rm N}^2} \frac{q_c}{(2\pi)^2 \, v_F}\int_{(\Delta \varepsilon/v_F)^2}^\infty \, dq^2 \frac{1}{\sqrt{q^2-(\Delta\varepsilon/v_F)^2}}\frac{1}{\xi ^{-2} (h,t) + q^2} \nonumber\\
&=&\frac{\pi^2 T}{2 (2\pi)^2} \frac{q_c \, p_F}{ N(0) \, v_F}\frac{1}{p_F \, \xi_{\rm N}^2} \Big(\xi ^{-2} (h,t) + (\Delta \varepsilon/v_F)^2 \Big)^{-1/2} .
\end{eqnarray}

\end{document}